\documentclass[twocolumn,showpacs,amsmath,amssymb,aps,prx,footinbib,floatfix,superscriptaddress,longbibliography]{revtex4-1}
\usepackage{graphicx}
\usepackage{braket}
\usepackage{savesym}
\usepackage{amsfonts}
\usepackage{bm}
\usepackage{color}
\usepackage{subfigure}
\usepackage{wasysym}
\usepackage{upgreek}
\usepackage{float}
\usepackage{dsfont}
\usepackage{mathtools}
\usepackage{multirow}
\usepackage[usenames,dvipsnames,svgnames,table]{xcolor}
\usepackage{hyperref}
\usepackage{hhline} 
\usepackage{multirow} 
\usepackage[normalem]{ulem}

\usepackage{makecell}
\usepackage[english]{babel}

\hypersetup{
	colorlinks=true,       
	linkcolor=VioletRed,  
    citecolor=JungleGreen,        
    filecolor=Orchid,      
    urlcolor=black           
}
\begin{document}

\title{Recurrent neural network wave functions for Rydberg atom arrays on kagome lattice} 

\author{Mohamed Hibat-Allah}
\email{mhibatallah@uwaterloo.ca}
\affiliation{Department of Applied Mathematics, University of Waterloo, Waterloo, ON N2L 3G1, Canada}
\affiliation{Perimeter Institute for Theoretical Physics, 31 Caroline St N, Waterloo, ON N2L 2Y5, Canada}
\affiliation{Department of Physics and Astronomy, University of Waterloo, Ontario, N2L 3G1, Canada}
\affiliation{Vector Institute,  Toronto,  Ontario,  M5G 0C6,  Canada}

\author{Ejaaz Merali}
\affiliation{Perimeter Institute for Theoretical Physics, 31 Caroline St N, Waterloo, ON N2L 2Y5, Canada}
\affiliation{Department of Physics and Astronomy, University of Waterloo, Ontario, N2L 3G1, Canada}

\author{Giacomo Torlai}
\affiliation{AWS Center for Quantum Computing, Pasadena, CA, USA}

\author{Roger G Melko}
\affiliation{Perimeter Institute for Theoretical Physics, 31 Caroline St N, Waterloo, ON N2L 2Y5, Canada}
\affiliation{Department of Physics and Astronomy, University of Waterloo, Ontario, N2L 3G1, Canada}

\author{Juan Carrasquilla}
\affiliation{Institute for Theoretical Physics, ETH Zürich, 8093, Switzerland}
\affiliation{Vector Institute,  Toronto,  Ontario,  M5G 0C6,  Canada}
\affiliation{Department of Physics and Astronomy, University of Waterloo, Ontario, N2L 3G1, Canada}

\date{\today}

\begin{abstract}

Rydberg atom array experiments have demonstrated the ability to act as powerful quantum simulators, preparing strongly-correlated phases of matter which are challenging to study for conventional computer simulations. A key direction has been the implementation of interactions on frustrated geometries, in an effort to prepare exotic many-body states such as spin liquids and glasses. In this paper, we apply two-dimensional recurrent neural network (RNN) wave functions to study the ground states of Rydberg atom arrays on the kagome lattice. We implement an annealing scheme to find the RNN variational parameters in regions of the phase diagram where exotic phases may occur, corresponding to rough optimization landscapes. For Rydberg atom array Hamiltonians studied previously on the kagome lattice, our RNN ground states show no evidence of exotic spin liquid or emergent glassy behavior. In the latter case, we argue that the presence of a non-zero Edwards-Anderson order parameter is an artifact of the long autocorrelations times experienced with quantum Monte Carlo (QMC) simulations, and we show that autocorrelations can be systematically reduced by increasing numerical effort. This result emphasizes the utility of autoregressive models, such as RNNs, in conjunction with QMC, to explore Rydberg atom array physics on frustrated lattices and beyond. 
\end{abstract}

\maketitle

\section{Introduction}

Rydberg atom arrays have emerged as a rich playground for quantum simulation of many-body problems~\cite{browaeys_many-body_2020}. A key property of these arrays is their high degree of programmability, which enables the realization of multiple Hamiltonians on different lattice geometries and parameter ranges. This programmability facilitates the simulation of a wide array of phases of matter~\cite{Ebadi2021, wurtz2023aquila} and enables the solution to challenging combinatorial optimization problems~\cite{wurtz2023aquila, Ebadi22,Nguyen_2023}. Remarkably, the preparation of spin liquid phases---disordered phases of matter characterized by the presence of anyonic excitations, topological invariants, and long-range entanglement---has been demonstrated in programmable Rydberg arrays, potentially serving as building blocks of future generation of fault-tolerant qubits~\cite{Dennis_2002, kitaevAnyonsExactlySolved2006,kitaev2009topological}.

Recent numerical studies have investigated the physics of the ground state of Rydberg atom arrays in different lattice geometries, in particular in one~\cite{PhysRevA.98.023614} and two spatial dimensions in various geometries~\cite{Samajdar_2020, kalinowski2021bulk, Li_22, RydbergHarvard2021, RubyDMRG2021, PhysRevE.106.034121, Kornjaca2023}. In lattices such as ruby and honeycomb lattices, strong numerical evidence favours the existence of a spin liquid phase in agreement with experiments~\cite{RubyDMRG2021, Kornjaca2023}. Another recent example is the kagome lattice, where Density Matrix Renormalization Group (DMRG)~\cite{DMRG1992, DMRG2011} studies provided evidence that Rydberg atom arrays host a liquid-like regime~\cite{RydbergHarvard2021}, while Quantum Monte Carlo (QMC) simulations predicted the existence of a spin glass phase~\cite{Yan2023}. These systems display frustration arising from lattice geometry and Hamiltonian interactions, leading to the existence of a large number of quantum states with nearly degenerate energies but markedly different properties. This makes it computationally difficult to accurately approximate the ground state of these systems. 

Here we focus on applying recurrent neural network (RNNs) wave functions ~\cite{RNNWF, roth2020iterative} to a Rydberg array of atoms on the kagome lattice. The effectiveness of RNNs and Transformer language models has already been demonstrated in Rydberg atom arrays on the square lattice~\cite{moss2023enhancing, sprague2023variational, Czischek_2022}. RNNs possess two key properties that make them particularly well-suited for studying frustrated systems. Firstly, their ability to perform exact sampling helps mitigate frustration-induced ergodicity issues in quantum Monte Carlo. Secondly, the ability to define them in any spatial dimension without incurring additional computational intractability helps address challenges faced by techniques like DMRG, such as the increased computational cost stemming from increased entanglement in higher dimensions~\cite{RNNWF, RNN_Symmetry_Annealing}.  

Our findings reveal that in the highly frustrated and highly entangled regimes of the system, the RNN predicts a paramagnetic phase without topological order, consistent with earlier QMC simulations~\cite{Yan2023}. However, in contrast to the QMC results in Ref.~\onlinecite{Yan2023}, the RNN suggests the absence of a spin-glass phase. Nevertheless, in agreement with QMC, our numerical simulations indicate the emergence of a rugged optimization landscape, necessitating more optimization steps and thermal-like fluctuations to mitigate local minima in the RNN's parameter landscape.

Overall, our results showcase the remarkable applicability and advantages of machine learning-based wave functions, particularly RNNs, in tackling challenging problems at the forefront of Rydberg atom array physics. These findings pave the way for further exploration of exotic phases and phenomena in highly frustrated quantum systems, harnessing the power of modern machine learning techniques to advance our understanding in this field.

\section{Methods}

We focus our attention on an array of neutral atoms
on the kagome lattice, interacting via laser excitation to atomic Rydberg states. We consider a lattice with periodic boundary conditions (PBC). The Hamiltonian of this system is given by~\cite{browaeys_many-body_2020,RydbergHarvard2021}:
\begin{align*}
    \hat{H} &= - \sum_{i = 1}^{N} \frac{\Omega}{2} \Big( \ket{g}_i \bra{r}_i + \ket{r}_i \bra{g}_i \Big) - \delta \sum_{i = 1}^{N} \ket{r}_i \bra{r}_i \\
    &+ \frac{1}{2} \sum_{i,j} V(|| \bm{x}_i - \bm{x}_j||) \ket{r}_i \bra{r}_i \otimes \ket{r}_j \bra{r}_j.
\end{align*}
Here $\ket{g}_i, \ket{r}_i$ are respectively the ground and excited states of the Rydberg atom $i$. $\Omega$ is the Rabi frequency and $\delta$ is the laser detuning. $V(R) = C/R^6$ is the repulsive potential due to the dipole-dipole interaction between Rydberg atoms, which is responsible for the blockade mechanism~\cite{browaeys_many-body_2020}. In practice, we define a blockade radius $R_b$ such that $V(R_b/a) = \Omega$, where $a$ is the distance between two neighbouring Rydberg atoms. Finally, we note that the sum over all possible pairs is truncated to a sum over neighbors separated by a distance cutoff $R_c = 2$ or $R_c = 4$. The choice $R_c = 2$ is taken to compare with the DMRG results reported in Ref.~\cite{RydbergHarvard2021} as well as with the QMC findings in Ref.~\cite{Yan2023}.

\subsection{Two dimensional RNNs}
\label{sec:2DRNN}

\begin{figure*}
    \centering
    \includegraphics[width =\linewidth]{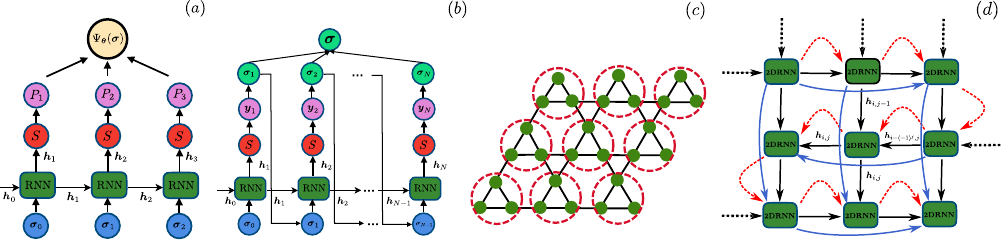}
    \caption{(a) An illustration of a positive RNN wave function. Each RNN cell receives an input $\bm{\sigma}_{n-1}$ and a hidden state $\bm{h}_{n-1}$ and outputs a new hidden state $\bm{h}_n$. This vector is taken as an input to the Softmax layer (denoted S) that computes the conditional probability $P_i$. (b) RNN autoregressive sampling scheme: after obtaining the probability vector $\bm{y}_i$ from the Softmax layer (S) in step $i$, we sample it to produce $\bm{\sigma}_i$. The latter is taken again as an input to the RNN along with the hidden state $\bm{h}_i$ to sample the following degree of freedom $\bm{\sigma}_{i+1}$. (c) Mapping of a Kagome lattice to a square lattice by embedding three atoms in a larger local Hilbert space. (d) A two-dimensional (2D) RNN with periodic boundary conditions for a $3 \times  3$ lattice for illustration purposes. A bulk RNN cell receives two hidden states $\bm{h}_{i,j-1}$ and $\bm{h}_{i-(-1)^j,j}$, as well as two input vectors $\bm{\sigma}_{i,j-1}$ and $\bm{\sigma}_{i-(-1)^j,j}$ (not shown) as illustrated by the black solid arrows. RNN cells at the boundary receive additional hidden states $\bm{h}_{i,j+1}$ and $\bm{h}_{i+(-1)^j,j}$, as well as two input vectors $\bm{\sigma}_{i,j+1}$ and $\bm{\sigma}_{i+(-1)^j,j}$ (not shown), as demonstrated by the blue curved and solid arrows. The sampling path is taken as a zigzag path, as demonstrated by the dashed red arrows. The initial memory states of the 2D RNN and the initial inputs are null vectors, as indicated by the dashed black arrows.}
    \label{fig:RNN}
\end{figure*}

The Rydberg Hamiltonian is stoquastic in nature~\cite{bravyi2015monte}, which implies that the ground-state wave function contains only positive amplitudes. This offers the opportunity to model the ground state with an RNN wave function with only positive amplitudes~\cite{RNNWF} which we adopt below. Complex extensions of RNN wave functions for non-stoquastic Hamiltonians have been explored in Refs.~\cite{RNNWF, roth2020iterative, RNN_Symmetry_Annealing}. To model a positive RNN wave function, we can express our ansatz in the computational basis as:
\begin{equation*}
   \Psi_{\bm{\theta}}(\bm{\sigma}) = \sqrt{p_{\bm{\theta}}(\bm{\sigma})},
\end{equation*}
such that $\bm{\theta}$ corresponds to the variational parameters of the ansatz $\ket{\Psi_{\bm{\theta}}}$, and $\bm{\sigma} = (\sigma_1, \sigma_2, \ldots, \sigma_N)$ is a configuration of the Rydberg atoms. The main advantage of using RNN wave functions is the possibility of estimating observables through autoregressive sampling, which allows obtaining uncorrelated samples by construction~\cite{RNNWF}. To do so, we model the joint probability $p_{\bm{\theta}}(\bm{\sigma})$ by constructing the conditionals $p_{\bm{\theta}}(\sigma_i | \sigma_{<i})$ by taking advantage of the probability chain rule
\begin{equation*}
    p_{\bm{\theta}}(\bm{ \sigma})= p_{\bm{\theta}}(\sigma_1)p_{\bm{\theta}}(\sigma_2|\sigma_1) \cdots p_{\bm{\theta}}(\sigma_N|\sigma_{N-1}, \dots, \sigma_2, \sigma_1).
\end{equation*}
These conditional probabilities are obtained through a Softmax layer as follows:
\begin{equation*}
    p_{\bm{\theta}}(\sigma_i | \sigma_{<i}) = \bm{y}_i \cdot \bm{\sigma}_i.
\end{equation*}
Here $\bm{y}_i = \text{Softmax}(U \bm{h}_{i} + \bm{c})$ where $U$ and $\bm{c}$ are, respectively, trainable weights and biases, and `Softmax' corresponds to the normalizing Softmax activation function. Additionally, the memory (hidden) state $\bm{h}_{\rm i}$ is obtained recursively as~\cite{lipton2015}:
\begin{equation}
    \bm{h}_{i} = f(W[\bm{h}_{i-1}; \bm{\sigma}_{i-1}] + \bm{b}),
    \label{eq:1DRNN}
\end{equation}
such that $[. ; .]$ is a concatenation of two vectors, while $\bm{\sigma}_{i-1}$ is a one-hot encoding of $\sigma_{i-1}$. These computations are illustrated in Fig.~\ref{fig:RNN}(a). $W$ and $\bm{b}$ are also trainable weights and biases, and $f$ is a user-defined activation function. 

By virtue of the `Softmax' activation function, the conditionals $p_{\bm{\theta}}(\sigma_i | \sigma_{<i})$ are normalized to one. This property implies that the RNN joint probability $p_{\bm{\theta}}$ is also normalized~\cite{RNNWF}. Furthermore, by sampling the conditionals $p_{\bm{\theta}}(\sigma_i | \sigma_{<i})$ sequentially, as illustrated in Fig.~\ref{fig:RNN}(b), we can extract exact samples from the joint RNN probability $p_{\bm{\theta}}$. An attractive property of this scheme is the possibility to efficiently generate uncorrelated samples from different modes present in $p_{\bm{\theta}}$, whereas traditional Metropolis sampling scheme may get stuck in only one mode.

The atom configurations of a Rydberg atom array on a kagome lattice can be seen as an $L \times L \times 3$ array of binary degrees of freedom where $L$ is the size of each side of the lattice. As illustrated in Fig.~\ref{fig:RNN}(c), we can map our kagome lattice with a local Hilbert space of $2$ to a square lattice with an enlarged Hilbert space of size $2^3 = 8$ which we can study using our two-dimensional (2D) RNN wave function~\cite{RNN_topological_order, sprague2023variational}.

To construct a 2D RNN ansatz that can handle PBC, we modify our RNN recursion in Eq.~\eqref{eq:1DRNN} to a two-dimensional recursion relation as:
\begin{align}
    \bm{h}_{i,j} &= f\! \Big(
    W[\text{Neighbours}(\bm{h}_{i,j}); \text{Neighbours}(\bm{\sigma}_{i,j})]
    +  \bm{b} \Big).
    \label{eq:2DRNN}
\end{align}
$\bm{h}_{i,j}$ is a memory state with two indices for each atom in the two-dimensional lattice. Here `Neighbours($\bm{\sigma}_{i,j}$)' returns a concatenation of the neighbours of $\bm{\sigma}_{i,j}$. The same observation goes for `Neighbours($\bm{h}_{i,j}$)'. These neighbours correspond to incoming vectors indicated by the black and blue arrows as illustrated in Fig.~\ref{fig:RNN}(d). More specifically, we define
\begin{align*}
\text{Neighbours}(\bm{h}_{i,j}) \equiv& \ [\bm{h}_{i-(-1)^j,j}; \bm{h}_{i,j-1}; \bm{0}; \bm{0}]
\end{align*}
on the bulk. On the boundaries, we take
\begin{align*}
\text{Neighbours}(\bm{h}_{i,j}) \equiv& \ [\bm{h}_{i-(-1)^j,j}; \bm{h}_{i,j-1}; \nonumber\\
&\bm{h}_{i+(-1)^j,j}; \bm{h}_{i,j+1}].
\end{align*}
Note that PBC on the indices is assumed. The additional inputs $\bm{\sigma}_{i+(-1)^j,j}$, $\bm{\sigma}_{i,j+1}$ and hidden states $\bm{h}_{i+(-1)^j,j}$, $\bm{h}_{i,j+1}$ allow to take PBC into account and to introduce correlations between degrees of freedom at the boundaries. During the autoregressive sampling procedure, the input and hidden vectors are initialized to a null vector if not previously defined to preserve the autoregressive nature of our scheme, as illustrated in Fig.~\ref{fig:RNN}(b). Also, note that the particular choice of the indices is motivated by the zigzag sampling path. In this study, we use an advanced version of 2D RNNs incorporating the gating mechanism as previously done in Refs.~\cite{RNN_Symmetry_Annealing, Vieijra2021, luo2021gauge}. More details can be found in Appendix.~\ref{app:GRU}. Finally, since $\bm{h}_{i,j}$ is a summary of the history of the generated $\sigma_{<i,j}$, it is used to compute the conditional probabilities as follows:
\begin{equation}
    p_{\bm{\theta}}(\sigma_{i,j}| \sigma_{<i,j}) = \text{Softmax}(U \bm{h}_{i,j} + \bm{c}) \cdot \bm{\sigma}_{i,j}.
    \label{eq:prob_softmax}
\end{equation}

\subsection{Supplementing RNN optimization with annealing}
\label{sec:annealing}
To reach the ground state of the Rydberg atoms array Hamiltonian on the kagome lattice, we minimize the energy expectation value $E_{\bm{\theta}} = \bra{\Psi_{\bm{\theta}}} \hat{H}\ket{\Psi_{\bm{\theta}}}$ using the Variational Monte Carlo (VMC) scheme~\cite{becca_sorella_2017} (see Appendix~\ref{app:VMC}). Due to the frustrated nature of the kagome lattice which can induce local minima in the VMC scheme, we leverage annealing with thermal-like fluctuations to mitigate local minima. This technique has been suggested and implemented in Refs.~\cite{roth2020iterative, VNA2021, RNN_Symmetry_Annealing, Roth2022,RNNOpt2023,RNN_topological_order}. In this case, we obtain a free-energy like cost function, defined as
\begin{equation}
    F_{\bm{\theta}}(n) = E_{\bm{\theta}} - T(n) S_{\rm classical} ( p_{\bm{\theta}} ),
    \label{eq:FreeEnergy}
\end{equation}
where $F_{\bm{\theta}}$ is a variational pseudo Free energy and $S_{\rm classical}$ is the classical Shannon entropy:
\begin{equation}
    S_{\rm classical} (p_{\bm{\theta}}) = - \sum_{\bm{\sigma}} p_{\bm{\theta}}(\bm{\sigma}) \log\left(p_{\bm{\theta}}(\bm{\sigma})\right).
    \label{eq:vnEntropy}
\end{equation}
The previous sum goes over all classical Rydberg configurations $\{\bm{\sigma}\}$ in the computational $z$-basis. Note that $S_{\rm classical}$ is a pseudo-entropy that can be efficiently estimated using our RNN wave function as opposed to the quantum von Neumann entropy. Additionally, $T(n)$ is a pseudo-temperature that is annealed from some initial value $T_0$ to zero as follows: $T(n) = T_0 (1-n/N_{\rm a})$ where $n \in [ 0, N_{\rm a} ]$ and $N_{\rm a}$ is the total number of annealing steps. We note that for each annealing step, we train our RNN for $5$ training steps. We present more details about the hyperparameters of our training scheme in Appendix.~\ref{app:hyperparams}.

\subsection{Topological entanglement entropy}
\label{sec:TEE}

To investigate the existence of a topological property in the Rydberg atom arrays on the kagome lattice, we compute the topological entanglement entropy (TEE)~\cite{Hamma2004,Hamma2005,LevinWen2006,KitaevPreskill2006, Hamma2009, TEE2011, TEE2017}. For a gapped phase of matter, where the area law is satisfied, the Renyi-$2$ entanglement entropy follows the scaling law $S_2(A) = aL - \gamma + \mathcal{O}(L^{-1})$, assuming $A$ and $B$ is partition of the system, $L$ is the size of the boundary between $A$ and $B$ and $S_2(A) \equiv - \log(\text{Tr}(\rho^2_A))$. In this case, $\gamma$ is the so-called TEE. In this paper, we use the swap trick with our RNN wave function ansatz~\cite{EE2010, RNNWF, EE2020} to calculate the second Renyi entropy $S_2$ to extract the TEE $\gamma$. 

\begin{figure}
    \centering
    \includegraphics[width =\linewidth]{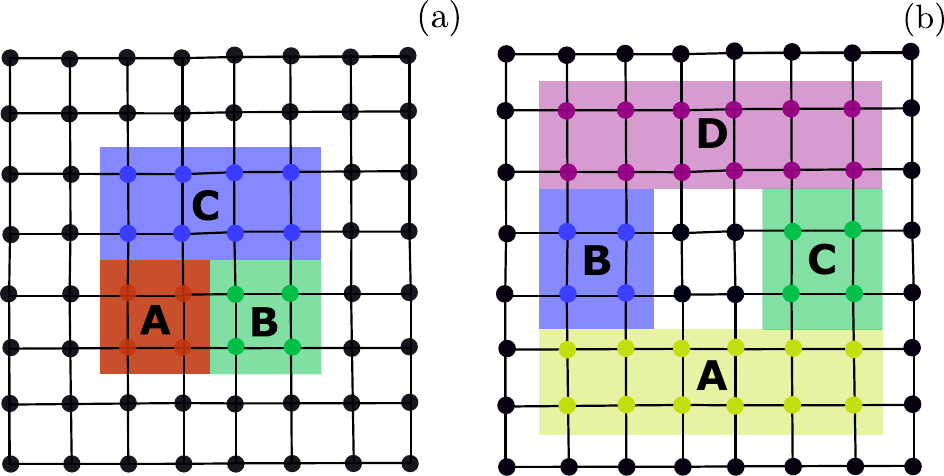}
    \caption{(a) A sketch of the parts $A$, $B$, and $C$ that we use for Kitaev-Preskill construction to compute the TEE. (b) Levin-Wen construction using the regions $A, B, C$ and $D$. For the Rydberg atoms Hamiltonian on a kagome lattice, each dot on the square lattice corresponds to a block of three binary degrees of freedom, as shown in Fig.~\ref{fig:RNN}(c).}
    \label{fig:TEE_constructions}
\end{figure}

We extract $\gamma$ using two different strategies, namely the Kitaev-Preskill construction~\cite{KitaevPreskill2006} and the Levin-Wen construction~\cite{LevinWen2006}, illustrated in Fig.~\ref{fig:TEE_constructions}.
The Kitaev-Preskill construction consists of choosing three subregions $A$, $B$, $C$ with geometries as shown in Fig.~\ref{fig:TEE_constructions}(a). The TEE can be then obtained by computing
\begin{align*}
    \gamma &= -S_2(A)-S_2(B)-S_2(C)+S_2(AB)\\
    &+S_2(AC)+S_2(BC) - S_2(ABC),
\end{align*}
where $S_2(A)$ is the second Renyi entropy of the subsystem $A$, and $AB$ is the union of $A$ and $B$ and similarly for the other terms. It is worth mentioning that finite size effects on $\gamma$ can be reduced by extrapolating the size of the subregions~\cite{KitaevPreskill2006, Furukawa2007}. Finally, note that this approach combined with RNN wave functions was successful in extracting a non-zero TEE on the toric code and the hard-core Bose-Hubbard model on the kagome lattice~\cite{RNN_topological_order}.

The Levin-Wen construction allows to extract the TEE $\gamma$ by constructing four different subsystems $A_1 = A \cup B \cup C \cup D,A_2 = A \cup C \cup D, A_3 = A \cup B \cup D$ and $A_4 = A \cup D$ as illustrated in Fig.~\ref{fig:TEE_constructions}(b) such that~\cite{TEE2011}:
\begin{equation*}
    \gamma = \frac{-S_2(A_1) + S_2(A_2) + S_2(A_3) - S_2(A_4)}{2}.
\end{equation*}
Note that finite size effects on $\gamma$ can be reduced by extrapolating the width and thickness of $A_1, A_2, A_3$ and $A_4$~\cite{TEE2011, Furukawa2007}.

Finally, we would like to highlight that our ability to study quantum systems with fully periodic boundary conditions is key to mitigating boundary effects, as opposed to cylinders used in DMRG~\cite{Stoudenmire2012,Cylinders2014}, which can introduce boundary effects in the TEE value~\cite{RydbergHarvard2021}.

\section{Results}
\label{sec:results}

\begin{figure*}
    \centering
    \includegraphics[width = 0.9\linewidth]{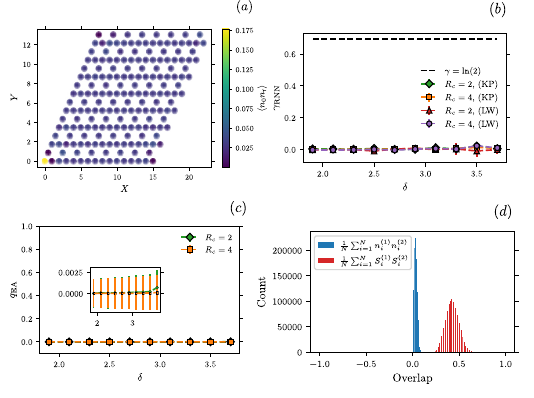}
    \caption{In these panels, we focus on the Blockade radius $R_b = 1.95$. (a) Plot of two point correlations $\langle n_{\bm{0}} n_{\bm{r}} \rangle$ with $\delta = 3.3$ and for a system size $N = 8 \times 8 \times 3$ and $R_c = 2$. (b) Plots of the topological entanglement entropy versus $\delta$ for two different values of the cutoff radius $R_c$, using the Levin-Wen (LW) construction and the Kitaev-Preskill (KP) construction, for $N = 8 \times 8 \times 3$. (c) A histogram of the Edwards-Anderson order parameter $q_{\text{EA}}$ defined in Eq.~\eqref{eq:EA_order} as a function of $\delta$ for $N = 8 \times 8 \times 3$. (d) A plot of the density overlaps $\frac{1}{N} \sum_{i=1}^{N} n_i^{(1)} n_i^{(2)}$ and the spin overlaps $\frac{1}{N} \sum_{i=1}^{N} S_i^{(1)} S_i^{(2)}$ at $\delta = 3.3$. Here $S_i = 2 n_i - 1$, and (1) and (2) are labels for two sets of samples obtained from our optimized RNN, that are aggregated from $10$ different training seeds, for $N = 6 \times 6 \times 3$ and $R_c = 2$. For each seed, we generate $2 \times 10^5$ independent samples and divide them into two sets. The error bars indicate the statistical uncertainty of one standard deviation, calculated across different samples. In this and other plots throughout the Article, error bars may be hidden if they are smaller than the symbol size.}
    \label{fig:RydbergRb195}
\end{figure*}

According to the RNN numerics, our results show that the ground state at $R_b = 1.95$ and $\delta = 3.3$, which is suggested to be in the spin-liquid phase according to Ref.~\cite{RydbergHarvard2021}, is rather a disordered state with no topological order. We first plot the correlations $\langle n_{\bm{0}} n_{\bm{r}} \rangle $ in Fig.~\ref{fig:RydbergRb195}(a). The results indicate that the extracted state has short-range correlations. To confirm the correctness of our variational implementation, we perform a sanity check and compare our ground state energies with QMC and DMRG as shown in Appendix~\ref{app:numerical_check}. We found a good agreement between our RNN energies and QMC as well as DMRG energies. Most importantly, we observe that our RNN results using only $d_h = 100$ are more accurate compared to DMRG with a bond dimension $\chi = 1000$ in the highly entangled regime at $R_b = 1.95$ and $\delta = 3.3$.

To investigate the existence of a spin liquid in this regime, we calculate the TEE $\gamma$ using the Kitaev-Preskill construction~\cite{KitaevPreskill2006} for a system size $L = 8$ (see Fig.~\ref{fig:TEE_constructions}(a)), and for different values of $\delta \in [2.0, 3.7]$ and $R_c = 2, 4$ at $R_b = 1.95$. We also do the same using the Levin-Wen construction~\cite{TEE2011} in Fig.~\ref{fig:TEE_constructions}(b). Our results, illustrated in Fig.~\ref{fig:RydbergRb195}(b) suggest that the TEE extracted by the RNN is consistent with zero and different from $\ln(2)$ within error bars. These results suggest the non-existence of a spin liquid within our settings and also suggest that the state we find in this regime is a disordered state. Our findings are further corroborated by a recent QMC study~\cite{Yan2023} and also by previous results in the literature suggesting that the paramagnetic `liquid' phase in Ising systems on the kagome lattice is not exotic~\cite{Nikolic2005,Moessner2001, Moessner_2000}. 

To address the finite-size scaling of the TEE, we compute the TEE at $R_b = 1.95, \delta = 3.3$ for $N = 10\times10\times3$ by pre-training from the RNN optimized at $L = 8$ and also for $L = 12$ by starting from the optimized parameters at $L = 10$. We follow a similar KP construction to Fig.~\ref{fig:TEE_constructions} of the regions $A,B$ and $C$ where the size of each subregion is given as $3\times 3\times 3$, $3\times 3\times 3$, and $6\times 3\times 3$ for $L = 6$ respectively, and $4\times 4\times 3$, $4\times 4\times 3$, and $8\times 4\times 3$ for $L = 8$. Our estimates $\gamma_{\rm RNN}= 0.10 \pm 0.26, -0.1 \pm 0.22$ for $L = 6$ and $L = 8$ respectively. These values, which are consistent with zero TEE, corroborate the absence of a $Z_2$ spin-liquid according to the RNN variational calculations. Note that the error bars can be systematically reduced by increasing the number of samples in the swap trick calculations~\cite{RNNWF, RNN_topological_order}.

In this QMC study~\cite{Yan2023}, it was suggested that the region, around $R_b = 1.95$ and the values of $\delta$ used in our study, contains an emergent spin-glass phase instead of a paramagnetic state. To verify this claim, we compute the Edwards-Anderson (EA) order parameters~\cite{EA_1975, Richards84}, defined as:
\begin{equation}
    q_{\text{EA}} = \frac{\sum_{i = 1}^{N} \langle n_i - \rho \rangle^2}{N \rho (1-\rho)},
    \label{eq:EA_order}
\end{equation}
where $N$ is the system size, $n_i$ is the occupation number of site $i$ and $\rho = (\sum_{i = 1}^{N} n_i)/N$. Deviations of this order parameter from zero values are signals of the existence of a spin-glass phase. In Fig.~\ref{fig:RydbergRb195}(c), we plot this order parameter as a function of $\delta$ with $R_c = 2,4$ and $R_b = 1.95$. We find that the values of the order parameter are consistent with zero, as opposed to the results of QMC in Ref.~\cite{Yan2023}. Furthermore, we report in Fig.~\ref{fig:RydbergRb195}(d) the density-density overlap $\frac{1}{N} \sum_{i=1}^{N} n_i^{(1)} n_i^{(2)}$ and the spin-spin overlap $\frac{1}{N} \sum_{i=1}^{N} S_i^{(1)} S_i^{(2)}$ between different RNN samples at $R_b = 1.95, \delta = 3.3$, and $R_c = 2$. Here labels (1) and (2) correspond to two independent sets of samples, which are obtained from optimized RNNs with $10$ different training seeds. The Gaussian nature of the overlap distribution in both representations is another indicator that there is no static signature of a spin-glass order~\cite{Castellani_2005}. 

The discrepancy in our results and previous QMC findings~\cite{Yan2023} could be related to emergent glassy dynamics in the QMC simulations, which results in very long auto-correlations times and thus in a non-ergodic behavior. To corroborate our findings, we run QMC simulations~\cite{merali2023stochastic}, based on Stochastic Series Expansion (SSE)~\cite{PhysRevB.43.5950, Sandvik_1999}, for larger inverse temperatures compared to Ref.~\cite{Yan2023}, namely for $\beta \geq 200$ and using $2.2 \times 10^6$ Monte Carlo samples. We find that the QMC prediction for the EA order parameter is given as $q^{\text{QMC}}_{\text{EA}} = 0.0000018(5)$ for $R_b = 1.95, \delta = 3.3$, a system size $8\times 8\times 3$, and for a radius cutoff $R_c = 2$. The previous result agrees very well with our RNN findings in Fig.~\ref{fig:RydbergRb195}(c). This result is also confirmed by the good agreement between the RNN energies and the QMC energies as shown in Appendix.~\ref{app:numerical_check}. Our findings are further supported by the results of Ref.~\cite{Yan2022}, which suggests the possibility of transition in a quantum dimer model between nematic to paramagnetic to staggered states. In conclusion, our numerical investigation suggests that the long auto-correlation time could be a limiting factor in the QMC results reported in Ref.~\cite{Yan2023}.

\begin{figure*}[htb]
    \centering
    \includegraphics[width = \linewidth]{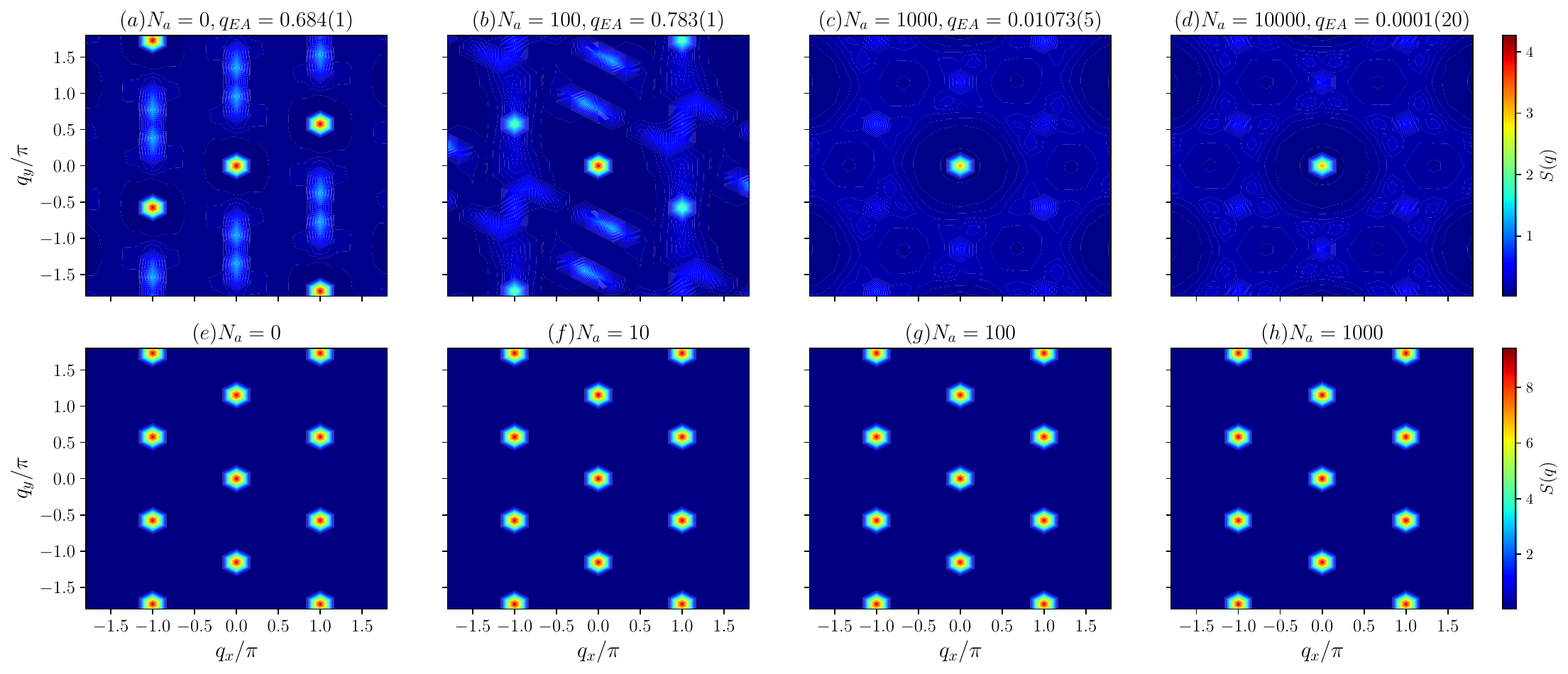}
    \caption{Plots of the structure factor for different values of the number of annealing steps $N_a$ at $R_b = 1.95, \delta = 3.3$ in panels (a-d) and at $R_b = 1.7, \delta = 3.3$ in panels (e-h). The color bars represent the magnitude of the structure factor $S(\bm{q})$. Additionally, these results correspond to a system size $N = 6 \times 6 \times 3$. Note that we observe convergence to a paramagnetic state without ordering peaks beyond $N_a = 1000$ for $R_b = 1.95, \delta = 3.3$. For $R_b = 1.7, \delta = 3.3$, we find that the nematic state is not affected by the choice of $N_a$ and can be obtained without a need for annealing.}
\label{fig:annealing_structurefactor}
\end{figure*}

We note that the emergence of a long autocorrelation time in QMC coincides with the emergence of a rugged optimization landscape, which in our simulations implies a longer number of annealing steps in our RNN simulations to achieve convergence. To demonstrate this point, we compute the structure factor
\begin{equation}
    S(\bm{q}) = \frac{1}{N} \sum_{i,j} \langle n_i n_j \rangle \ e^{\text{i} \bm{q}\cdot(\bm{x}_i - \bm{x}_j) }
\end{equation}
to extract the nature of the states obtained by our RNN ansatz and investigate their dependence on the number of annealing steps $N_a$. We expect the optimization landscape to be rougher as $N_a$ required to converge increases.
Figs.~\ref{fig:annealing_structurefactor}(a-d) at $R_b = 1.95$ and $\delta = 3.3$ show that the RNN finds different states for different numbers of annealing steps $N_a$, until it converges to a state without ordering peaks, i.e., the paramagnetic state. In contrast, the nematic state at $R_b = 1.7, \delta = 3.3$ can be reached without the need for annealing, as illustrated by the structure factors at different $N_a$ in Fig.~\ref{fig:annealing_structurefactor}(e-h). These observations suggest an emergent rugged optimization landscape when optimizing our ansatz in the highly entangled regime. Finally, to find the optimal number of annealing steps $N_a$ in the highly entangled regime, we note that we conduct a scaling study as shown in Appendix~\ref{app:annealing}.

To indicate the quality of our variational calculations, we use the v-score as a metric~\cite{v_score}, which we report in Appendix~\ref{app:v-score}. Furthermore, to investigate the effect of parameter sharing in the RNN, illustrated in Eqs.~\eqref{eq:2DRNN},~\eqref{eq:prob_softmax}, we optimized our RNN using site-dependent parameters at $R_b = 1.95, \delta = 3.3$ and for $L = 8$ and we find that both TEE and EA order parameters are consistent with zero within errors bars. This result confirms that parameter-sharing in our RNN ansatz does not bias our findings. More details are shared in Appendix~\ref{app:weight_sharing}.

Finally, we report a comparison between RNN and QMC runtimes in Appendix~\ref{app:time}. In particular, we find a significant speed-up when training RNNs on an A100 GPU and QMC on a single CPU. This speedup is enabled by GPU hardware in addition to the ability of the 2DRNN to do transfer learning from smaller lattices to larger ones~\cite{roth2020iterative, RNN_Symmetry_Annealing, RNN_topological_order,luo2021gauge, moss2025leveragingrecurrenceneuralnetwork, moss2025_triangular}.

\section{Conclusions and Outlooks}

In this paper, we demonstrate a successful application of recurrent neural network (RNN) wave functions to the task of investigating topological order on Rydberg atom arrays on kagome lattice. We use these architectures to estimate the second Renyi entropies using the swap trick~\cite{RNNWF}. The latter allows us to compute the TEEs using the Kitaev-Preskill~\cite{KitaevPreskill2006} and the Levin-Wen~\cite{LevinWen2006} constructions. Furthermore, with the possibility of handling periodic boundary conditions in RNNs, the boundary effects on the TEE are reduced compared to DMRG, which has challenges with boundary effects on cylinders~\cite{Cylinders2014}.

Our main finding, suggested by the two-dimensional RNN wave functions results, points out that Rydberg atom arrays on the kagome lattice do not establish a $Z_2$ spin liquid in the highly entangled regime. This observation is also consistent with previous QMC studies~\cite{Yan2023}. Our RNN numerics also suggest that the highly entangled region corresponds to a trivial paramagnetic state and that there is no signature for spin glass order as opposed to the observations outlined in Ref.~\cite{Yan2023}. We believe that the ability of RNNs to generate uncorrelated samples from a multimodal distribution is a crucial factor for our numerics to indicate the non-existence of the spin-glass phase. Additionally, we find that autocorrelation could be the main factor behind the spin glass phase observed in previous QMC simulations~\cite{Yan2023}. In particular, our QMC numerics with more numerical effort compared to Ref.~\cite{Yan2023} suggest the absence of a spin glass phase. Furthermore, supplementing RNNs with annealing turns out to be a valuable tool for mitigating local minima induced by the frustrated nature of the kagome lattice in the highly entangled regime. We highlight that advanced optimization schemes, such as minimal Stochastic Reconfiguration minSR~\cite{Chen_2024, Rende_2024}, is a potential avenue for enhancing the optimization of our 2DRNN ans\"atzes. Refs.~\cite{donatella_autoregressive, lange2024neural} reported that Stochastic Reconfiguration (SR) is not as effective as Adam optimizer when applied to RNN wave functions. As a result, we believe that reconciling SR and RNN wave function optimization is an interesting research direction that deserves a thorough study in the future. Furthermore, studying dynamic properties of Rydberg atoms arrays is another promising research direction that can corroborate our findings by performing time-evolution on our RNN ansatz~\cite{sinibaldi2024timedependentneuralgalerkinmethod, vandewalle2024manybodydynamicsexplicitlytimedependent}.

Finally, we note that our method can be generalized to study other systems with potential topological order, such as the Rydberg atom arrays on the Ruby lattice~\cite{RubyDMRG2021, RydbergSimulator2021, giudici2022dynamical}. One could also use quantum state tomography with RNNs~\cite{Carrasquilla2019} in a wide variety of quantum simulators and also combine data from QMC or quantum simulators with VMC to improve the variational results~\cite{Bennewitz2021NeuralEM, Czischek_2022, moss2023enhancing, Lange_2025}. We also believe in the potential of RNN wave functions ans\"atzes in the discovery of new phases of matter with topological order. Overall, these results highlight the promising future of RNN wave functions~\cite{RNNWF, roth2020iterative}, language-model based wave functions, and neural quantum states~\cite{Carleo2017} for investigating open questions and discovering new physics within the condensed matter community and beyond.

\section*{Code Availability}
Our code is made publicly available at ``\url{http://github.com/mhibatallah/RNNWavefunctions}''. The hyperparameters we use are given in Appendix~\ref{app:hyperparams}.

\section*{Data availability}
The data generated in this study is available from the corresponding author upon reasonable request.


\section*{Acknowledgments}
We would like to thank Subir Sachdev, Anders Sandvik, and Arun Paramekanti for their helpful and inspiring discussions. Our RNN implementation is based on Tensorflow~\cite{tensorflow2015-whitepaper} and NumPy~\cite{Harris2020}.
Computer simulations were made possible thanks to the Vector Institute computing cluster and the Digital Research Alliance of Canada cluster. We acknowledge support from Natural Sciences and Engineering Research Council of Canada (NSERC), the Shared Hierarchical Academic Research Computing Network (SHARCNET), Compute Canada, and the Canadian Institute for Advanced Research (CIFAR) AI chair program. This work is not related to the research being performed at AWS. Research at Perimeter Institute is supported in part by the Government of Canada through the Department of Innovation, Science and Economic Development and by the Province of Ontario through the Ministry of Colleges and Universities. This research was supported in part by grant NSF PHY-2309135 to the Kavli Institute for Theoretical Physics (KITP).
\appendix 
\section{Two dimensional periodic gated RNNs}
\label{app:GRU}
In this Appendix, we share more details about our 2D gated RNN wave function implementation for periodic systems, which is used in this study to target the ground states of the Rydberg atom arrays on the kagome lattice. If we define
\begin{align*}
    \bm{h'}_{i,j} &= [\bm{h}_{i-(-1)^j,j} ; \bm{h}_{i,j-1}; \bm{h}_{i+(-1)^j,j}; \bm{h}_{i,j+1}], \\
    \bm{\sigma'}_{i,j} &= [\bm{\sigma}_{i-(-1)^j,j} ; \bm{\sigma}_{i,j-1}; \bm{\sigma}_{i+(-1)^j,j}; \bm{\sigma}_{i,j+1}],
\end{align*}
then our gated 2D RNN wave function ansatz is based on the following recursion relations:
\begin{align*}
    \bm{h}_{i,j} &= \tanh \! \Big(
    W[\bm{\sigma'}_{i,j}; \bm{h'}_{i,j}]
    +  \bm{b} \Big), \\
   \bm{u}_{i,j} &= \text{sigmoid}\! \Big(
    W_g[\bm{\sigma'}_{i,j}; \bm{h'}_{i,j}]
    +  \bm{b}_g \Big), \\
    \bm{h}_{i,j} &= \bm{u}_{i,j} \odot \bm{\tilde{h}}_{i,j} + (1-\bm{u}_{i,j}) \odot (U_g \bm{h'}_{i,j}).
\end{align*}
A hidden state $\bm{h}_{i,j}$ can be obtained by combining a candidate state $\bm{\tilde{h}}_{i,j}$ and the neighbouring hidden states $\bm{h}_{i-1,j}, \bm{h}_{i,j-1}, \bm{h}_{i+1,j}, \bm{h}_{i,j+1}$. The update gate $\bm{u}_{i,j}$ determines how much of the candidate hidden state $\bm{\tilde{h}}_{i,j}$ will be taken into account and how much of the neighboring states will be considered. With this combination, it is possible to mitigate some limitations of the vanishing gradient problems~\cite{zhou2016minimal,shen2019mutual}. The weight matrices $W, W_g, U_g$ and the biases $b, b_g$ are variational parameters of our RNN ansatz in addition to the Softmax layer parameters in Eq.~\eqref{eq:prob_softmax}. Note that we choose the size of the hidden state $\bm{h}_{i,j}$, which we denote as $d_h$, before optimizing our ansatz parameters. We note that the choice of the gated 2DRNN is motivated by its superiority compared to the non-gated 2DRNN on the task of finding the ground state of the 2D Heisenberg model~\cite{RNN_Symmetry_Annealing}. 

Since we use an enlarged local Hilbert space with three atoms at each recursion step, the size of the Softmax layer output is defined as $2^3$. Additionally, each input $\bm{\sigma}_{i,j}$ is defined as a concatenation of the one-hot encoding of each of the three atoms. This means that $\bm{\sigma}_{i,j}$ is a six-dimensional vector. 

\section{Variational monte carlo (VMC)}
\label{app:VMC}
To optimize the energy expectation value of our RNN wave function $\Psi_{\bm{\theta}}$, we use the Variational Monte Carlo (VMC) scheme, which consists of using importance sampling to estimate the energy expectation value $E_{\bm{\theta}} = \bra{\Psi_{\bm{\theta}}} \hat{H} \ket{\Psi_{\bm{\theta}}}$ as follows~\cite{becca_sorella_2017,RNNWF}:
\begin{equation*}
    E_{\bm{\theta}} = \frac{1}{M}\sum_{i=1}^{M} E_{\text{loc}}(\bm{\sigma^{(i)}}),
\end{equation*}
where the local energies $E_{\text{loc}}$ are defined as
\begin{equation*}
    E_{\text{loc}}(\bm{\sigma}) = \sum_{\bm{\sigma'}} H_{\bm{\sigma} \bm{\sigma'}} \frac{\Psi_{\bm{\theta}}(\bm{\sigma'})}{\Psi_{\bm{\theta}}(\bm{\sigma})}.
\end{equation*}
Here the configurations $\{\bm{\sigma^{(i)}}\}_{i = 1}^{M}$ are sampled from our ansatz using autoregressive sampling. The choice of $M$ is a hyperparameter that can be tuned. Similarly, the gradients can be estimated as
\begin{equation*}
    \partial_{\bm{\theta}} E_{\bm{\theta}} = \frac{1}{M}\sum_{i=1}^{M} \partial_{\bm{\theta}} \log \left( \Psi_{\bm{\theta}}^{*}(\bm{\sigma^{(i)}}) \right) \left ( E_{\text{loc}}(\bm{\sigma^{(i)}}) - E_{\bm{\theta}} \right).
\end{equation*}
Subtracting the mean energy $E_{\bm{\theta}}$ is helpful to achieve convergence as it reduces the variance of the gradients without biasing its expectation value~\cite{RNNWF, RNN_topological_order}. The gradient descent steps are performed using the Adam optimizer~\cite{AdamPaper}. Similarly to the stochastic energy estimation, we can implement a similar procedure for the estimation of the variational pseudo-free energy $F_{\bm \theta}$ in Eq.~\eqref{eq:FreeEnergy}. Ref.~\cite{VNA2021} provides more details in the supplementary information.

\section{Hyperparameters}
\label{app:hyperparams}

For all models studied in this paper, we note that for each annealing step, we perform $N_{\rm train} = 5$ gradient steps. Concerning the learning rate $\eta$, we choose $\eta = 10^{-3}$ during the warm-up phase and the annealing phase and switch to a learning rate $\eta = 10^{-4}$ in the convergence phase. To train RNN on the $8 \times 8 \times 3$ lattices, we pre-train using the optimized RNN on the $6 \times 6 \times 3$ lattice without using annealing, since the RNN is expected to start from a variational energy that is close to the ground state energy in the new system size.

In Tab.~\ref{tab:hyperparams}, we provide further details about the hyperparameters we choose in our study for the different models. For the estimation of the RNN energy, we use $2 \times 10^5$ independent configurations. We also use $M = 2 \times 10^6$ independent samples for the estimation of the entanglement entropy along with their error bars. For the estimation of the TEE using Kitaev-Preskill or Levin-Wen constructions, we use the expression of the standard deviation of the sum of independent random variables to estimate the one standard deviation on $\gamma_{\rm RNN}$.

\begin{table*}[p]
    \centering
    \scriptsize
    \begin{tabular}{|c|c|c|}\hline
       Figures & Parameter & Value \\\hline   
      \multirow{8}{*}{Fig.~\ref{fig:RydbergRb195} ($N = 6 \times 6 \times 3$)} & Number of memory units & $d_h = 100$ \\
            & Number of training samples & $M = 500$ \\
            & Initial pseudo-temperature & $T_0 = 2$ \\
            & Number of warm-up steps
             & $N_{w} = 1000$ \\    
            & Number of annealing steps
             & $N_{a} = 10000$ \\    
             & Number of convergence steps
             & $N_{\text{conv}} = 10000$ \\ 
              & Number of samples for TEE estimation
             & $M = 10^7$ \\
             & Number of samples for $q_{\text{EA}}$ estimation
             & $M = 2 \times 10^5$ \\ 
           \hline            
      \multirow{6}{*}{Fig.~\ref{fig:RydbergRb195} ($N = 8 \times 8 \times 3$)} & Number of memory units & $d_h = 100$ \\
            & Number of training samples & $M = 500$ \\
            & Initial pseudo-temperature & $T_0 = 0$ \\
            & Number of warm-up steps
             & $N_{w} = 0$ \\  
            & Number of annealing steps
             & $N_{a} = 0$ (pre-trained from $N = 6 \times 6 \times 3$) \\    
             & Number of training steps
             & $N_{\text{train}} = 10000$ \\      
       \hline            
      \multirow{5}{*}{Fig.~\ref{fig:annealing_structurefactor} ($N = 6 \times 6 \times 3$)} & Number of memory units & $d_h = 100$ \\
            & Number of training samples & $M = 500$ \\
            & Initial pseudo-temperature & $T_0 = 2$ \\ 
            & Number of warm-up steps
             & $N_{w} = 1000$ \\  
             & Number of convergence steps
             & $N_{\text{conv}} = 10000$ \\
              & Number of sample for two-point correlations estimation & $M = 2 \times 10^5$ \\
       \hline              
      \multirow{8}{*}{Fig.~\ref{fig:annealing_effect} ($N = 6 \times 6 \times 3$)} & Number of memory units & $d_h = 100$ \\
            & Number of training samples & $M = 500$ \\
            & Initial pseudo-temperature & $T_0 = 2$ \\ 
            & Number of warm-up steps
             & $N_{w} = 1000$ \\ 
             & Number of convergence steps
             & $N_{\text{conv}} = 10000$ \\ 
              & Number of samples for energy and density estimation
             & $M = 2 \times 10^5$ \\
              & Number of samples for $S_2$ estimation
             & $M = 2 \times 10^7$ \\ 
       \hline             
    \end{tabular}
    \caption{A summary of the hyperparameters used to obtain the results reported in this paper.} 
    \label{tab:hyperparams}
\end{table*}

\section{Comparison with QMC and DMRG}
\label{app:numerical_check}

\begin{figure*}
    \centering
    \includegraphics[width=0.8\linewidth]{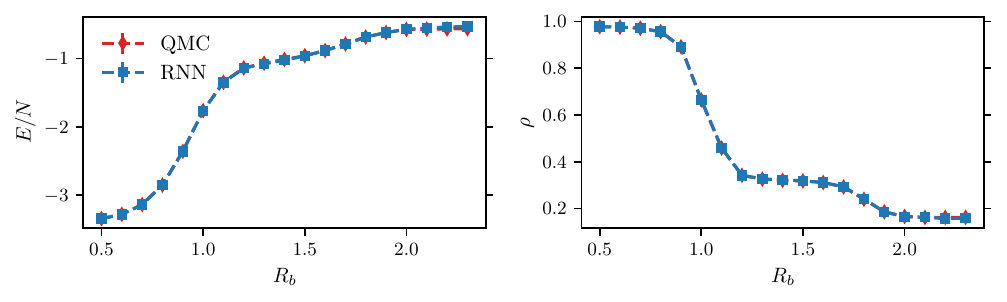}
    \caption{A comparison between RNN and QMC energies per site $E/N$ (left panel) and densities $\rho$ (right panel) for multiple values of $R_b$, fixed $\delta = 3.3$ and $R_c = 2$ on a system size $8 \times 8 \times 3$.}
    \label{fig:RNN_QMC_compare}
\end{figure*}

In this appendix, we conduct a comparison between RNNs, QMC, and DMRG. First of all, we compare RNNs (with $N_a = 4000$) and QMC from $R_b = 0.5$ to $2.3$ at fixed $\delta = 3.3$ at a system size $8 \times 8 \times 3$ and a cutoff radius $R_c = 2$. We report our results in Fig.~\ref{fig:RNN_QMC_compare}, where we find a very good agreement between our QMC and RNN energies per site $E/N$ and densities $\rho$. In particular, we observe that $\rho$ plateaus close to three different values, $1, 1/3$, and $1/6$ respectively. The latter are consistent with previous findings~\cite{Yan2023}. In particular, $\rho \approx 1/3$ corresponds to the nematic phase and $\rho \approx 1/6$ to the highly-entangled disordered phase.

\begin{figure}
    \centering
\includegraphics[width=\linewidth]{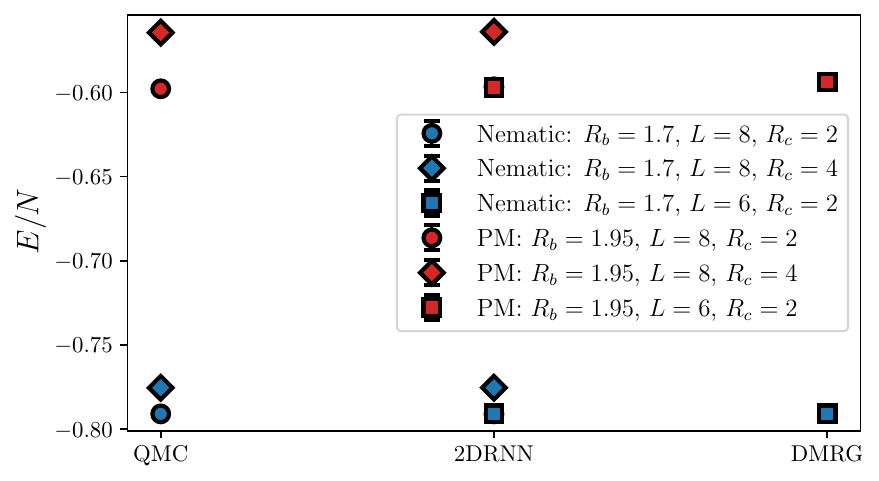}
\caption{Comparison between our RNN ansatz, QMC, and DMRG within the Nematic and paramagnetic (PM) phases and for different system sizes $L$ and cutoff radii $R_c$. Numerical values are reported in Tabs.~\ref{tab:QMCcomparison},~~\ref{tab:DMRGcomparison}.}
\label{fig:appendixD_fig}
\end{figure}

In Tab.~\ref{tab:QMCcomparison}, we show an additional comparison between RNN's (with $N_a = 10000)$ and QMC's energies per site $E/N$ for a system size $8 \times 8 \times 3$ and for a detuning $\delta = 3.3$ and at the blockade radii $R_b = 1.7, 1.95$. These points correspond to the nematic and disordered phases, respectively. We note that our RNN-based ansatz provides energies with a relative error of less than $0.2\%$ compared to the QMC energies. The QMC simulations we run for Rydberg atom arrays are introduced in Ref.~\cite{merali2023stochastic}. We use a finite-temperature QMC scheme run at several different values of $\beta$ until we observe convergence to the ground state. For each $\beta$, five independent simulations are taken, and the convergence of observables is observed at $\beta = 200$. Thus, to compute observables, we treat simulations with $\beta \geq 200$ as additional independent chains, giving us a total of 25 independent Markov chains at each parameter point. Each chain is allowed to warm-up for $10^4$ steps, after which $10^6$ sequential measurements were taken. With respect to the computation of the Edwards-Anderson order parameter, $q_{\rm EA}$, we note that the analysis given in Ref.~\cite{Yan2023} can give different results in the case of imperfect sampling. Ref.~\cite{Yan2023} computes the order parameter independently for each Markov chain and then averages the results. This procedure can produce different results as each chain will only explore a subset of the QMC configuration space due to the presence of frustrated interactions. As a result, each chain's estimate of the one-point function can be biased. Since $q_{\text{EA}}$ is a non-linear function of the one-point function, we must first aggregate the one-point functions generated by each Markov chain, and then compute $q_{\rm EA}$. Lastly, to compute an estimate of the error in $q_{\rm EA}$, we must account for auto-correlations and non-linearity simultaneously. This step is done by combining jackknife resampling with a binning procedure. To deal with auto-correlations, we first compute the one-point function on sequential ``bins'' of data; we found a bin size of $10^4$ to be sufficient, giving $100$ bins for each chain. Thus, we can consider each bin's one-point function to be nearly uncorrelated, allowing us to apply the jackknife resampling procedure to these approximately independent bins. To visualize our additional numerical comparisons with DMRG and QMC on fully periodic boundary conditions, we report our results in Fig.~\ref{fig:appendixD_fig}.

To compare our RNN wave function ($d_h = 100$) results with DMRG, we perform DMRG simulations using PastaQ~\cite{pastaq} and ITensor~\cite{Fishman_2022} to further check the consistency of our RNN energies. In Tab.~\ref{tab:DMRGcomparison}, we compare with DMRG using periodic boundary conditions. In the nematic phase $R_b = 1.7, \delta = 3.3$, we find an excellent match of the energies. For the disordered phase at $R_b = 1.95, \delta = 3.3$, our RNN energies are lower within an error of about $1\%$ and with orders of magnitude fewer parameters compared to DMRG. Furthermore, we choose the YC12 geometry used in Ref.~\cite{RydbergHarvard2021}. We optimize our 2DRNN wave function ($d_h = 60$) at $R_b = 1.95$ and $\delta = 3.5$. Our estimated energy is $-151.959(8)$ which is within 0.3\% error compared to the DMRG energy provided in Ref.~\cite{RydbergHarvard2021}.

\begin{table*}[]
\begin{tabular}{|c|c|c|c|c|}
\hline
 & QMC ($R_c = 2$) & 2DRNN ($R_c = 2$) & QMC ($R_c = 4$) & 2DRNN ($R_c = 4$) \\ \hline
$R_b = 1.7, \delta = 3.3$   & -0.790975(14) & -0.790964(5) &  -0.77546(1)  & -0.775412(4)  \\ \hline
$R_b = 1.95, \delta = 3.3$ & -0.59785(1)  & -0.59666(2) & -0.56445(1) & -0.56401(3) \\ \hline
\end{tabular}
\caption{A table of the energies per site obtained by QMC and 2DRNN for a system size $8 \times 8 \times 3$ with fully periodic boundary conditions and for different values of the cutoff radius $R_c$. The error bars in parentheses correspond to the one-standard deviation uncertainty in the QMC and RNN energy estimates.}
\label{tab:QMCcomparison}
\end{table*}

\begin{table*}[]
\begin{tabular}{|c|c|c|}
\hline
Rydberg parameters & DMRG ($R_c = 2$) & 2DRNN ($R_c = 2$) \\ \hline
$R_b = 1.7, \delta = 3.3$ & -0.790957  & -0.790934(7) \\ \hline
$R_b = 1.95, \delta = 3.3$ & -0.593828  & -0.59716(2) \\ \hline
\end{tabular}
\caption{A table comparing DMRG (with bond dimension $\chi = 110$ for $R_b = 1.7$ and $\chi = 1000$ for $R_b = 1.95$) and the 2DRNN ansatz in terms of the energies per site for a system size $6 \times 6 \times 3$ with fully periodic boundary conditions and for a cutoff radius $R_c = 2$ and $\delta = 3.3$. Note that a DMRG run on $R_b = 1.95, \delta = 3.3$ for $\chi = 1500$ was not successful with a memory allocation of $120$ GB. To run our RNN simulations, an $80$ GB memory allocation was sufficient. The error bars in the parentheses correspond to the one-standard deviation uncertainty on the RNN energy.}
\label{tab:DMRGcomparison}
\end{table*}

\section{Annealing and local minima}
\label{app:annealing}

\begin{figure*}
    \centering
    \includegraphics[width = \linewidth]{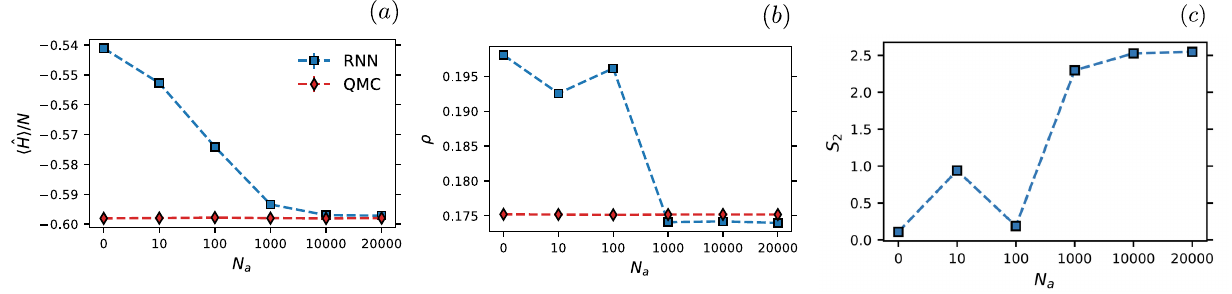}
    \caption{Plots of the energy density $\langle \hat{H} \rangle/N$, density $\rho = (\sum_{i = 1}^{N} n_i)/N$ and the second Renyi entropy $S_2$ as function of the number of annealing steps $N_a$ for a system size $N = 6 \times 6 \times 3$ and $R_b = 1.95$, $\delta = 3.3$ and $R_c = 2$. We observe a saturation of all these observables in the asymptotic limit, which justifies the use of $N_a = 10000$ in our numerical simulations in Fig.~\ref{fig:RydbergRb195}. For comparison, we also report the QMC data obtained using annealing for the energy per site and density.}
    \label{fig:annealing_effect}
\end{figure*}

In Fig.~\ref{fig:annealing_effect}, we demonstrate the importance of incorporating annealing in the training of RNNs applied to Rydberg atom arrays on the kagome lattice. These experiments are carried out in the highly entangled regime at $R_b = 1.95$ and $\delta = 3.3$. In panel (a), we observe that the ground state energy improves with more annealing steps $N_a$. We also highlight that the density saturates close to $1/6$, which raises the possibility of an odd quantum spin liquid~\cite{Yan2023, RydbergHarvard2021}. However, it is not a sufficient condition. This phase is found to be the paramagnetic state according to our RNN numerics. We also outline a saturation of the second Renyi entropy to a large value in the asymptotic limit of $N_a$. All these numerics suggest that $N_a = 10000$ is a good choice to converge our RNN training.

Furthermore, we note that we implement annealing in QMC and report our results in Fig.~\ref{fig:annealing_effect}, where we find that more annealing steps do not systematically improve QMC energies as opposed to annealing with RNNs. Finally, we highlight the strong agreement between RNN and QMC data for large $N_a$.

\section{V-Score values}
\label{app:v-score}
In Tab.~\ref{tab:vscores}, we report the v-score values~\cite{v_score} for the variational calculations of our 2D RNN for $L = 6, 8$ and at the hard and easy points ($R_b = 1.95$ and $1.7$ respectively) for fixed $\delta = 3.3$. The v-score is computed as:
\begin{equation}
    \text{v-score} = \frac{N \text{Var}(\hat{H})}{(\langle \hat{H} \rangle-E_{\infty})^2},
\end{equation}
where $\langle \hat{H} \rangle,\text{Var}(\hat{H})$ are respectively the expectation values and the variances of the Hamiltonian $\hat{H}$ in our optimized 2DRNN ansatz. $E_{\infty}$ is a reference energy at infinite temperature, which is computed here with a Monte Carlo estimate using $10^5$ samples~\cite{v_score}.

The estimated v-scores are a good indication of the quality of a variational calculation~\cite{v_score}. We find that our values of the v-score in Tab.~\ref{tab:vscores} reflect the difficulty of the highly entangled paramagnetic phase at $R_b = 1.95, \delta = 3.3$ compared to the nematic phase at $R_b = 1.7, \delta = 3.3$. 

Within our setup, the values of $E_{\infty}$ are about one order to magnitude bigger than $\langle \hat{H} \rangle$. In this case, the denominator of the v-score is dominated by $E_{\infty}$. To avoid this issue, we compute another metric: the energy variance per spin $\text{Var}(\hat{H})/N$. These values, reported in Tab.~\ref{tab:vscores}, provide a clearer distinction between the quality of $R_b = 1.95, \delta = 3.3$ variational calculation compared to $R_b = 1.7, \delta = 3.3$, where there is close to one order of magnitude difference in terms of the energy variances per spin. 

\begin{table*}[]
\begin{tabular}{|c|c|c|c|}
\hline
$N$ & Rydberg parameters & V-score & Energy variance per spin \\ \hline
\multirow{2}{*}{$6\times6\times3$} 
& $R_b = 1.7, \delta = 3.3$ & $8.12 \times 10^{-6}$ & $1.16 \times 10^{-3}$ \\ 
& $R_b = 1.95, \delta = 3.3$ & $1.16\times 10^{-5}$ & $9.14 \times 10^{-3}$ \\ \hline
\multirow{2}{*}{$8\times8\times3$} 
& $R_b = 1.7, \delta = 3.3$ & $6.03\times 10^{-6}$ & $8.59 \times 10^{-4}$\\ 
& $R_b = 1.95, \delta = 3.3$ & $1.10\times 10^{-5}$ & $8.70 \times 10^{-3}$ \\ \hline
\end{tabular}
\caption{A table of V-score values and energy variances for different system sizes $N$ and different blockade radii $R_b$. These scores are obtained using the 2DRNN wave function trained using the hyperparameters reported in Tab.~\ref{tab:hyperparams} for Fig.~\ref{fig:RydbergRb195}.}
\label{tab:vscores}
\end{table*}

\section{Parameters sharing}
\label{app:weight_sharing}

The RNN recursion relation in Eq.~\eqref{eq:2DRNN} assumes the use of the same parameters across different sites, which is consistent with the translation invariance of the Rydberg atoms Hamiltonian. This parameter sharing property allows to iteratively train the RNN on larger system sizes by pre-training from previous system sizes.

In order to investigate the inductive bias of using weight sharing in our ansatz. We optimize our 2DRNN with no-parameter sharing, where the two-dimensional recursion relation is given as:
\begin{align}
    \bm{h}_{i,j} &= f\! \Big(
    W_{ij}[\text{Neighbours}(\bm{h}_{i,j}); \text{Neighbours}(\bm{\sigma}_{i,j})]
    +  \bm{b}
    _{ij}\Big),
    \label{eq:2DRNN_noweightsharing}
\end{align}
using position-dependent weights and biases $W_{ij}, \bm{b}_{ij}$. We also do the same for the Softmax layer, in Eq.~\eqref{eq:prob_softmax}, as follows:
\begin{equation}
    p_{\bm{\theta}}(\sigma_{i,j}| \sigma_{<i,j}) = \text{Softmax}(U_{ij} \bm{h}_{i,j} + \bm{c}_{ij}) \cdot \bm{\sigma}_{i,j},
    \label{eq:prob_softmax_noweightsharing}
\end{equation}
with site-dependent weights and biases $U_{ij}, \bm{c}_{ij}$.

We optimize at 2DRNN with site-dependent parameters for a system size $N = 8 \times 8 \times 3$ with $R_b = 1.95, \delta = 3.3$, and we report our results in Tab.~\ref{tab:NWScomparion}, where we find that the RNN results are consistent with zero $q_{\rm EA}$ and zero $\gamma_{\rm RNN}$ as predicted by the 2DRNN with weight sharing. We also find that the energy per-spin $E/N$ for the 2DRNN with weight sharing is lower compared to the 2DRNN with non-weight sharing, even though the latter has $N$ more parameters. This result highlights the importance of embedding inductive bias of the Hamiltonian in the wave function ansatz to improve variational calculations.

\begin{table*}[]
\begin{tabular}{|c|c|c|}
\hline
 & 2DRNN-NWS ($R_c = 2$) & 2DRNN ($R_c = 2$) \\ \hline
$E/N$ & -0.59447(2)  & -0.59666(2) \\\hline
$q_{\rm EA}$ & 0.001416(6)  & 0.000205(1) \\ \hline$\gamma_{\rm RNN}$ & -0.006(6)  & -0.003(20) \\ \hline
\end{tabular}
\caption{A table comparing the 2DRNN results with weight sharing with the 2DRNN with no-weight sharing (2DRNN-NWS). Here we use a system size $N = 8 \times 8 \times 3$ and $R_b = 1.95, \delta = 3.3$ for the comparison.}
\label{tab:NWScomparion}
\end{table*}

\section{Simulations runtime}
\label{app:time}

To provide an idea of the simulation running time, we share a table of the times it takes to run one RNN simulation on $6\times 6\times 3$, $8\times 8\times 3$ lattices given the hyperparameters shared in Tab.~\ref{tab:hyperparams}. Our simulations are originally run on a P100 GPU with 12GB of memory. Our simulation data, in Tab.~\ref{tab:time}, demonstrates that we can get our RNN trained for about a day and 4 hours on a P100 GPU for $L = 6$ and $L = 8$, respectively. Note that we used pre-training in the $L=8$ simulation, from the $L = 6$ simulation, to speed up training. We also showcase that this time can be further reduced using the A100 GPUs. Tab.~\ref{tab:time} shows that our runtime is significantly reduced to less than $3.5$ and $1.6$ hours for both $L = 6, 8$. This runtime is expected to be further reduced with more advanced GPU resources. We also report the average runtime for our QMC simulations in Tab.~\ref{tab:time} for different values of $R_b$ and at fixed $\delta = 3.3$. Note that QMC run times are explicitly dependent on the Hamiltonian parameters. Our comparison confirms that RNN training times are significantly lower compared to QMC runs on a single CPU.

\begin{table*}
\centering
\begin{tabular}{|c|c|c|c|}
\hline
L & RNN (P100) & RNN (A100) & QMC (CPU)  \\
\hline
6 & 26.567 h & \textbf{3.460 h} &  48.751 h \\
8 & 4.074 h & \textbf{1.537 h} & 70.422 h \\
\hline
\end{tabular}
\caption{Runtime (in hours) of RNNs simulation on P100 and A100 GPUs for different values of $L$ (we use $N_{\rm a} = 10000$ for $L = 6$ and $N_{\rm conv} = 10000$ converge steps for $L = 8$). We also report the average QMC runtime on a single CPU for different values of $R_b = 1.95$ and at fixed $\delta = 3.3$. All of these simulations are conducted with a cutoff radius $R_c = 2$. The lowest values are highlighted in bold.}
\label{tab:time}
\end{table*}

\clearpage
\bibliography{Biblio}

\begin{thebibliography}{78}%
\makeatletter
\providecommand \@ifxundefined [1]{%
 \@ifx{#1\undefined}
}%
\providecommand \@ifnum [1]{%
 \ifnum #1\expandafter \@firstoftwo
 \else \expandafter \@secondoftwo
 \fi
}%
\providecommand \@ifx [1]{%
 \ifx #1\expandafter \@firstoftwo
 \else \expandafter \@secondoftwo
 \fi
}%
\providecommand \natexlab [1]{#1}%
\providecommand \enquote  [1]{``#1''}%
\providecommand \bibnamefont  [1]{#1}%
\providecommand \bibfnamefont [1]{#1}%
\providecommand \citenamefont [1]{#1}%
\providecommand \href@noop [0]{\@secondoftwo}%
\providecommand \href [0]{\begingroup \@sanitize@url \@href}%
\providecommand \@href[1]{\@@startlink{#1}\@@href}%
\providecommand \@@href[1]{\endgroup#1\@@endlink}%
\providecommand \@sanitize@url [0]{\catcode `\\12\catcode `\$12\catcode `\&12\catcode `\#12\catcode `\^12\catcode `\_12\catcode `\%12\relax}%
\providecommand \@@startlink[1]{}%
\providecommand \@@endlink[0]{}%
\providecommand \url  [0]{\begingroup\@sanitize@url \@url }%
\providecommand \@url [1]{\endgroup\@href {#1}{\urlprefix }}%
\providecommand \urlprefix  [0]{URL }%
\providecommand \Eprint [0]{\href }%
\providecommand \doibase [0]{http://dx.doi.org/}%
\providecommand \selectlanguage [0]{\@gobble}%
\providecommand \bibinfo  [0]{\@secondoftwo}%
\providecommand \bibfield  [0]{\@secondoftwo}%
\providecommand \translation [1]{[#1]}%
\providecommand \BibitemOpen [0]{}%
\providecommand \bibitemStop [0]{}%
\providecommand \bibitemNoStop [0]{.\EOS\space}%
\providecommand \EOS [0]{\spacefactor3000\relax}%
\providecommand \BibitemShut  [1]{\csname bibitem#1\endcsname}%
\let\auto@bib@innerbib\@empty
\bibitem [{\citenamefont {Browaeys}\ and\ \citenamefont {Lahaye}(2020)}]{browaeys_many-body_2020}%
  \BibitemOpen
  \bibfield  {author} {\bibinfo {author} {\bibfnamefont {Antoine}\ \bibnamefont {Browaeys}}\ and\ \bibinfo {author} {\bibfnamefont {Thierry}\ \bibnamefont {Lahaye}},\ }\bibfield  {title} {\enquote {\bibinfo {title} {Many-body physics with individually controlled {Rydberg} atoms},}\ }\href {\doibase 10.1038/s41567-019-0733-z} {\bibfield  {journal} {\bibinfo  {journal} {Nature Physics}\ }\textbf {\bibinfo {volume} {16}},\ \bibinfo {pages} {132--142} (\bibinfo {year} {2020})}\BibitemShut {NoStop}%
\bibitem [{\citenamefont {Ebadi}\ \emph {et~al.}(2021)\citenamefont {Ebadi}, \citenamefont {Wang}, \citenamefont {Levine}, \citenamefont {Keesling}, \citenamefont {Semeghini}, \citenamefont {Omran}, \citenamefont {Bluvstein}, \citenamefont {Samajdar}, \citenamefont {Pichler}, \citenamefont {Ho}, \citenamefont {Choi}, \citenamefont {Sachdev}, \citenamefont {Greiner}, \citenamefont {Vuleti{\'{c}}},\ and\ \citenamefont {Lukin}}]{Ebadi2021}%
  \BibitemOpen
  \bibfield  {author} {\bibinfo {author} {\bibfnamefont {Sepehr}\ \bibnamefont {Ebadi}}, \bibinfo {author} {\bibfnamefont {Tout~T.}\ \bibnamefont {Wang}}, \bibinfo {author} {\bibfnamefont {Harry}\ \bibnamefont {Levine}}, \bibinfo {author} {\bibfnamefont {Alexander}\ \bibnamefont {Keesling}}, \bibinfo {author} {\bibfnamefont {Giulia}\ \bibnamefont {Semeghini}}, \bibinfo {author} {\bibfnamefont {Ahmed}\ \bibnamefont {Omran}}, \bibinfo {author} {\bibfnamefont {Dolev}\ \bibnamefont {Bluvstein}}, \bibinfo {author} {\bibfnamefont {Rhine}\ \bibnamefont {Samajdar}}, \bibinfo {author} {\bibfnamefont {Hannes}\ \bibnamefont {Pichler}}, \bibinfo {author} {\bibfnamefont {Wen~Wei}\ \bibnamefont {Ho}}, \bibinfo {author} {\bibfnamefont {Soonwon}\ \bibnamefont {Choi}}, \bibinfo {author} {\bibfnamefont {Subir}\ \bibnamefont {Sachdev}}, \bibinfo {author} {\bibfnamefont {Markus}\ \bibnamefont {Greiner}}, \bibinfo {author} {\bibfnamefont {Vladan}\ \bibnamefont {Vuleti{\'{c}}}}, \ and\ \bibinfo {author} {\bibfnamefont
  {Mikhail~D.}\ \bibnamefont {Lukin}},\ }\bibfield  {title} {\enquote {\bibinfo {title} {Quantum phases of matter on a 256-atom programmable quantum simulator},}\ }\href {\doibase 10.1038/s41586-021-03582-4} {\bibfield  {journal} {\bibinfo  {journal} {Nature}\ }\textbf {\bibinfo {volume} {595}},\ \bibinfo {pages} {227--232} (\bibinfo {year} {2021})}\BibitemShut {NoStop}%
\bibitem [{\citenamefont {Wurtz}\ \emph {et~al.}(2023)\citenamefont {Wurtz}, \citenamefont {Bylinskii}, \citenamefont {Braverman}, \citenamefont {Amato-Grill}, \citenamefont {Cantu}, \citenamefont {Huber}, \citenamefont {Lukin}, \citenamefont {Liu}, \citenamefont {Weinberg}, \citenamefont {Long}, \citenamefont {Wang}, \citenamefont {Gemelke},\ and\ \citenamefont {Keesling}}]{wurtz2023aquila}%
  \BibitemOpen
  \bibfield  {author} {\bibinfo {author} {\bibfnamefont {Jonathan}\ \bibnamefont {Wurtz}}, \bibinfo {author} {\bibfnamefont {Alexei}\ \bibnamefont {Bylinskii}}, \bibinfo {author} {\bibfnamefont {Boris}\ \bibnamefont {Braverman}}, \bibinfo {author} {\bibfnamefont {Jesse}\ \bibnamefont {Amato-Grill}}, \bibinfo {author} {\bibfnamefont {Sergio~H.}\ \bibnamefont {Cantu}}, \bibinfo {author} {\bibfnamefont {Florian}\ \bibnamefont {Huber}}, \bibinfo {author} {\bibfnamefont {Alexander}\ \bibnamefont {Lukin}}, \bibinfo {author} {\bibfnamefont {Fangli}\ \bibnamefont {Liu}}, \bibinfo {author} {\bibfnamefont {Phillip}\ \bibnamefont {Weinberg}}, \bibinfo {author} {\bibfnamefont {John}\ \bibnamefont {Long}}, \bibinfo {author} {\bibfnamefont {Sheng-Tao}\ \bibnamefont {Wang}}, \bibinfo {author} {\bibfnamefont {Nathan}\ \bibnamefont {Gemelke}}, \ and\ \bibinfo {author} {\bibfnamefont {Alexander}\ \bibnamefont {Keesling}},\ }\href@noop {} {\enquote {\bibinfo {title} {Aquila: Quera's 256-qubit neutral-atom quantum computer},}\ }
  (\bibinfo {year} {2023}),\ \Eprint {http://arxiv.org/abs/2306.11727} {arXiv:2306.11727 [quant-ph]} \BibitemShut {NoStop}%
\bibitem [{\citenamefont {Ebadi}\ \emph {et~al.}(2022)\citenamefont {Ebadi}, \citenamefont {Keesling}, \citenamefont {Cain}, \citenamefont {Wang}, \citenamefont {Levine}, \citenamefont {Bluvstein}, \citenamefont {Semeghini}, \citenamefont {Omran}, \citenamefont {Liu}, \citenamefont {Samajdar}, \citenamefont {Luo}, \citenamefont {Nash}, \citenamefont {Gao}, \citenamefont {Barak}, \citenamefont {Farhi}, \citenamefont {Sachdev}, \citenamefont {Gemelke}, \citenamefont {Zhou}, \citenamefont {Choi}, \citenamefont {Pichler}, \citenamefont {Wang}, \citenamefont {Greiner}, \citenamefont {Vuletić},\ and\ \citenamefont {Lukin}}]{Ebadi22}%
  \BibitemOpen
  \bibfield  {author} {\bibinfo {author} {\bibfnamefont {S.}~\bibnamefont {Ebadi}}, \bibinfo {author} {\bibfnamefont {A.}~\bibnamefont {Keesling}}, \bibinfo {author} {\bibfnamefont {M.}~\bibnamefont {Cain}}, \bibinfo {author} {\bibfnamefont {T.~T.}\ \bibnamefont {Wang}}, \bibinfo {author} {\bibfnamefont {H.}~\bibnamefont {Levine}}, \bibinfo {author} {\bibfnamefont {D.}~\bibnamefont {Bluvstein}}, \bibinfo {author} {\bibfnamefont {G.}~\bibnamefont {Semeghini}}, \bibinfo {author} {\bibfnamefont {A.}~\bibnamefont {Omran}}, \bibinfo {author} {\bibfnamefont {J.-G.}\ \bibnamefont {Liu}}, \bibinfo {author} {\bibfnamefont {R.}~\bibnamefont {Samajdar}}, \bibinfo {author} {\bibfnamefont {X.-Z.}\ \bibnamefont {Luo}}, \bibinfo {author} {\bibfnamefont {B.}~\bibnamefont {Nash}}, \bibinfo {author} {\bibfnamefont {X.}~\bibnamefont {Gao}}, \bibinfo {author} {\bibfnamefont {B.}~\bibnamefont {Barak}}, \bibinfo {author} {\bibfnamefont {E.}~\bibnamefont {Farhi}}, \bibinfo {author} {\bibfnamefont {S.}~\bibnamefont {Sachdev}},
  \bibinfo {author} {\bibfnamefont {N.}~\bibnamefont {Gemelke}}, \bibinfo {author} {\bibfnamefont {L.}~\bibnamefont {Zhou}}, \bibinfo {author} {\bibfnamefont {S.}~\bibnamefont {Choi}}, \bibinfo {author} {\bibfnamefont {H.}~\bibnamefont {Pichler}}, \bibinfo {author} {\bibfnamefont {S.-T.}\ \bibnamefont {Wang}}, \bibinfo {author} {\bibfnamefont {M.}~\bibnamefont {Greiner}}, \bibinfo {author} {\bibfnamefont {V.}~\bibnamefont {Vuletić}}, \ and\ \bibinfo {author} {\bibfnamefont {M.~D.}\ \bibnamefont {Lukin}},\ }\bibfield  {title} {\enquote {\bibinfo {title} {Quantum optimization of maximum independent set using rydberg atom arrays},}\ }\href {\doibase 10.1126/science.abo6587} {\bibfield  {journal} {\bibinfo  {journal} {Science}\ }\textbf {\bibinfo {volume} {376}},\ \bibinfo {pages} {1209--1215} (\bibinfo {year} {2022})},\ \Eprint {http://arxiv.org/abs/https://www.science.org/doi/pdf/10.1126/science.abo6587} {https://www.science.org/doi/pdf/10.1126/science.abo6587} \BibitemShut {NoStop}%
\bibitem [{\citenamefont {Nguyen}\ \emph {et~al.}(2023)\citenamefont {Nguyen}, \citenamefont {Liu}, \citenamefont {Wurtz}, \citenamefont {Lukin}, \citenamefont {Wang},\ and\ \citenamefont {Pichler}}]{Nguyen_2023}%
  \BibitemOpen
  \bibfield  {author} {\bibinfo {author} {\bibfnamefont {Minh-Thi}\ \bibnamefont {Nguyen}}, \bibinfo {author} {\bibfnamefont {Jin-Guo}\ \bibnamefont {Liu}}, \bibinfo {author} {\bibfnamefont {Jonathan}\ \bibnamefont {Wurtz}}, \bibinfo {author} {\bibfnamefont {Mikhail~D.}\ \bibnamefont {Lukin}}, \bibinfo {author} {\bibfnamefont {Sheng-Tao}\ \bibnamefont {Wang}}, \ and\ \bibinfo {author} {\bibfnamefont {Hannes}\ \bibnamefont {Pichler}},\ }\bibfield  {title} {\enquote {\bibinfo {title} {Quantum optimization with arbitrary connectivity using rydberg atom arrays},}\ }\href {\doibase 10.1103/prxquantum.4.010316} {\bibfield  {journal} {\bibinfo  {journal} {PRX Quantum}\ }\textbf {\bibinfo {volume} {4}} (\bibinfo {year} {2023}),\ 10.1103/prxquantum.4.010316}\BibitemShut {NoStop}%
\bibitem [{\citenamefont {Dennis}\ \emph {et~al.}(2002)\citenamefont {Dennis}, \citenamefont {Kitaev}, \citenamefont {Landahl},\ and\ \citenamefont {Preskill}}]{Dennis_2002}%
  \BibitemOpen
  \bibfield  {author} {\bibinfo {author} {\bibfnamefont {Eric}\ \bibnamefont {Dennis}}, \bibinfo {author} {\bibfnamefont {Alexei}\ \bibnamefont {Kitaev}}, \bibinfo {author} {\bibfnamefont {Andrew}\ \bibnamefont {Landahl}}, \ and\ \bibinfo {author} {\bibfnamefont {John}\ \bibnamefont {Preskill}},\ }\bibfield  {title} {\enquote {\bibinfo {title} {Topological quantum memory},}\ }\href {\doibase 10.1063/1.1499754} {\bibfield  {journal} {\bibinfo  {journal} {Journal of Mathematical Physics}\ }\textbf {\bibinfo {volume} {43}},\ \bibinfo {pages} {4452--4505} (\bibinfo {year} {2002})}\BibitemShut {NoStop}%
\bibitem [{\citenamefont {Kitaev}(2006)}]{kitaevAnyonsExactlySolved2006}%
  \BibitemOpen
  \bibfield  {author} {\bibinfo {author} {\bibfnamefont {Alexei}\ \bibnamefont {Kitaev}},\ }\bibfield  {title} {\enquote {\bibinfo {title} {Anyons in an exactly solved model and beyond},}\ }\href {\doibase 10.1016/j.aop.2005.10.005} {\bibfield  {journal} {\bibinfo  {journal} {Annals of Physics}\ }\bibinfo {series} {January {{Special Issue}}},\ \textbf {\bibinfo {volume} {321}},\ \bibinfo {pages} {2--111} (\bibinfo {year} {2006})}\BibitemShut {NoStop}%
\bibitem [{\citenamefont {Kitaev}\ and\ \citenamefont {Laumann}(2009)}]{kitaev2009topological}%
  \BibitemOpen
  \bibfield  {author} {\bibinfo {author} {\bibfnamefont {Alexei}\ \bibnamefont {Kitaev}}\ and\ \bibinfo {author} {\bibfnamefont {Chris}\ \bibnamefont {Laumann}},\ }\href@noop {} {\enquote {\bibinfo {title} {Topological phases and quantum computation},}\ } (\bibinfo {year} {2009}),\ \Eprint {http://arxiv.org/abs/0904.2771} {arXiv:0904.2771 [cond-mat.mes-hall]} \BibitemShut {NoStop}%
\bibitem [{\citenamefont {Samajdar}\ \emph {et~al.}(2018)\citenamefont {Samajdar}, \citenamefont {Choi}, \citenamefont {Pichler}, \citenamefont {Lukin},\ and\ \citenamefont {Sachdev}}]{PhysRevA.98.023614}%
  \BibitemOpen
  \bibfield  {author} {\bibinfo {author} {\bibfnamefont {Rhine}\ \bibnamefont {Samajdar}}, \bibinfo {author} {\bibfnamefont {Soonwon}\ \bibnamefont {Choi}}, \bibinfo {author} {\bibfnamefont {Hannes}\ \bibnamefont {Pichler}}, \bibinfo {author} {\bibfnamefont {Mikhail~D.}\ \bibnamefont {Lukin}}, \ and\ \bibinfo {author} {\bibfnamefont {Subir}\ \bibnamefont {Sachdev}},\ }\bibfield  {title} {\enquote {\bibinfo {title} {Numerical study of the chiral ${\mathbb{z}}_{3}$ quantum phase transition in one spatial dimension},}\ }\href {\doibase 10.1103/PhysRevA.98.023614} {\bibfield  {journal} {\bibinfo  {journal} {Phys. Rev. A}\ }\textbf {\bibinfo {volume} {98}},\ \bibinfo {pages} {023614} (\bibinfo {year} {2018})}\BibitemShut {NoStop}%
\bibitem [{\citenamefont {Samajdar}\ \emph {et~al.}(2020)\citenamefont {Samajdar}, \citenamefont {Ho}, \citenamefont {Pichler}, \citenamefont {Lukin},\ and\ \citenamefont {Sachdev}}]{Samajdar_2020}%
  \BibitemOpen
  \bibfield  {author} {\bibinfo {author} {\bibfnamefont {Rhine}\ \bibnamefont {Samajdar}}, \bibinfo {author} {\bibfnamefont {Wen~Wei}\ \bibnamefont {Ho}}, \bibinfo {author} {\bibfnamefont {Hannes}\ \bibnamefont {Pichler}}, \bibinfo {author} {\bibfnamefont {Mikhail~D.}\ \bibnamefont {Lukin}}, \ and\ \bibinfo {author} {\bibfnamefont {Subir}\ \bibnamefont {Sachdev}},\ }\bibfield  {title} {\enquote {\bibinfo {title} {Complex density wave orders and quantum phase transitions in a model of square-lattice rydberg atom arrays},}\ }\href {\doibase 10.1103/physrevlett.124.103601} {\bibfield  {journal} {\bibinfo  {journal} {Physical Review Letters}\ }\textbf {\bibinfo {volume} {124}} (\bibinfo {year} {2020}),\ 10.1103/physrevlett.124.103601}\BibitemShut {NoStop}%
\bibitem [{\citenamefont {Kalinowski}\ \emph {et~al.}(2022)\citenamefont {Kalinowski}, \citenamefont {Samajdar}, \citenamefont {Melko}, \citenamefont {Lukin}, \citenamefont {Sachdev},\ and\ \citenamefont {Choi}}]{kalinowski2021bulk}%
  \BibitemOpen
  \bibfield  {author} {\bibinfo {author} {\bibfnamefont {Marcin}\ \bibnamefont {Kalinowski}}, \bibinfo {author} {\bibfnamefont {Rhine}\ \bibnamefont {Samajdar}}, \bibinfo {author} {\bibfnamefont {Roger~G.}\ \bibnamefont {Melko}}, \bibinfo {author} {\bibfnamefont {Mikhail~D.}\ \bibnamefont {Lukin}}, \bibinfo {author} {\bibfnamefont {Subir}\ \bibnamefont {Sachdev}}, \ and\ \bibinfo {author} {\bibfnamefont {Soonwon}\ \bibnamefont {Choi}},\ }\bibfield  {title} {\enquote {\bibinfo {title} {Bulk and boundary quantum phase transitions in a square rydberg atom array},}\ }\href {\doibase 10.1103/PhysRevB.105.174417} {\bibfield  {journal} {\bibinfo  {journal} {Phys. Rev. B}\ }\textbf {\bibinfo {volume} {105}},\ \bibinfo {pages} {174417} (\bibinfo {year} {2022})}\BibitemShut {NoStop}%
\bibitem [{\citenamefont {Li}\ \emph {et~al.}(2022)\citenamefont {Li}, \citenamefont {Yang},\ and\ \citenamefont {Xu}}]{Li_22}%
  \BibitemOpen
  \bibfield  {author} {\bibinfo {author} {\bibfnamefont {Chang-Xiao}\ \bibnamefont {Li}}, \bibinfo {author} {\bibfnamefont {Sheng}\ \bibnamefont {Yang}}, \ and\ \bibinfo {author} {\bibfnamefont {Jing-Bo}\ \bibnamefont {Xu}},\ }\bibfield  {title} {\enquote {\bibinfo {title} {Quantum phases of rydberg atoms on a frustrated triangular-lattice array},}\ }\href {\doibase 10.1364/OL.450855} {\bibfield  {journal} {\bibinfo  {journal} {Opt. Lett.}\ }\textbf {\bibinfo {volume} {47}},\ \bibinfo {pages} {1093--1096} (\bibinfo {year} {2022})}\BibitemShut {NoStop}%
\bibitem [{\citenamefont {Samajdar}\ \emph {et~al.}(2021)\citenamefont {Samajdar}, \citenamefont {Ho}, \citenamefont {Pichler}, \citenamefont {Lukin},\ and\ \citenamefont {Sachdev}}]{RydbergHarvard2021}%
  \BibitemOpen
  \bibfield  {author} {\bibinfo {author} {\bibfnamefont {Rhine}\ \bibnamefont {Samajdar}}, \bibinfo {author} {\bibfnamefont {Wen~Wei}\ \bibnamefont {Ho}}, \bibinfo {author} {\bibfnamefont {Hannes}\ \bibnamefont {Pichler}}, \bibinfo {author} {\bibfnamefont {Mikhail~D.}\ \bibnamefont {Lukin}}, \ and\ \bibinfo {author} {\bibfnamefont {Subir}\ \bibnamefont {Sachdev}},\ }\bibfield  {title} {\enquote {\bibinfo {title} {Quantum phases of rydberg atoms on a kagome lattice},}\ }\href {\doibase 10.1073/pnas.2015785118} {\bibfield  {journal} {\bibinfo  {journal} {Proceedings of the National Academy of Sciences}\ }\textbf {\bibinfo {volume} {118}},\ \bibinfo {pages} {e2015785118} (\bibinfo {year} {2021})}\BibitemShut {NoStop}%
\bibitem [{\citenamefont {Verresen}\ \emph {et~al.}(2021)\citenamefont {Verresen}, \citenamefont {Lukin},\ and\ \citenamefont {Vishwanath}}]{RubyDMRG2021}%
  \BibitemOpen
  \bibfield  {author} {\bibinfo {author} {\bibfnamefont {Ruben}\ \bibnamefont {Verresen}}, \bibinfo {author} {\bibfnamefont {Mikhail~D.}\ \bibnamefont {Lukin}}, \ and\ \bibinfo {author} {\bibfnamefont {Ashvin}\ \bibnamefont {Vishwanath}},\ }\bibfield  {title} {\enquote {\bibinfo {title} {Prediction of toric code topological order from rydberg blockade},}\ }\href {\doibase 10.1103/physrevx.11.031005} {\bibfield  {journal} {\bibinfo  {journal} {Physical Review X}\ }\textbf {\bibinfo {volume} {11}} (\bibinfo {year} {2021}),\ 10.1103/physrevx.11.031005}\BibitemShut {NoStop}%
\bibitem [{\citenamefont {Yang}\ and\ \citenamefont {Xu}(2022)}]{PhysRevE.106.034121}%
  \BibitemOpen
  \bibfield  {author} {\bibinfo {author} {\bibfnamefont {Sheng}\ \bibnamefont {Yang}}\ and\ \bibinfo {author} {\bibfnamefont {Jing-Bo}\ \bibnamefont {Xu}},\ }\bibfield  {title} {\enquote {\bibinfo {title} {Density-wave-ordered phases of rydberg atoms on a honeycomb lattice},}\ }\href {\doibase 10.1103/PhysRevE.106.034121} {\bibfield  {journal} {\bibinfo  {journal} {Phys. Rev. E}\ }\textbf {\bibinfo {volume} {106}},\ \bibinfo {pages} {034121} (\bibinfo {year} {2022})}\BibitemShut {NoStop}%
\bibitem [{\citenamefont {Kornja{\v{c}}a}\ \emph {et~al.}(2023)\citenamefont {Kornja{\v{c}}a}, \citenamefont {Samajdar}, \citenamefont {Macr{\`i}}, \citenamefont {Gemelke}, \citenamefont {Wang},\ and\ \citenamefont {Liu}}]{Kornjaca2023}%
  \BibitemOpen
  \bibfield  {author} {\bibinfo {author} {\bibfnamefont {Milan}\ \bibnamefont {Kornja{\v{c}}a}}, \bibinfo {author} {\bibfnamefont {Rhine}\ \bibnamefont {Samajdar}}, \bibinfo {author} {\bibfnamefont {Tommaso}\ \bibnamefont {Macr{\`i}}}, \bibinfo {author} {\bibfnamefont {Nathan}\ \bibnamefont {Gemelke}}, \bibinfo {author} {\bibfnamefont {Sheng-Tao}\ \bibnamefont {Wang}}, \ and\ \bibinfo {author} {\bibfnamefont {Fangli}\ \bibnamefont {Liu}},\ }\bibfield  {title} {\enquote {\bibinfo {title} {Trimer quantum spin liquid in a honeycomb array of rydberg atoms},}\ }\href {\doibase 10.1038/s42005-023-01470-z} {\bibfield  {journal} {\bibinfo  {journal} {Communications Physics}\ }\textbf {\bibinfo {volume} {6}},\ \bibinfo {pages} {358} (\bibinfo {year} {2023})}\BibitemShut {NoStop}%
\bibitem [{\citenamefont {White}(1992)}]{DMRG1992}%
  \BibitemOpen
  \bibfield  {author} {\bibinfo {author} {\bibfnamefont {Steven~R.}\ \bibnamefont {White}},\ }\bibfield  {title} {\enquote {\bibinfo {title} {Density matrix formulation for quantum renormalization groups},}\ }\href {\doibase 10.1103/PhysRevLett.69.2863} {\bibfield  {journal} {\bibinfo  {journal} {Phys. Rev. Lett.}\ }\textbf {\bibinfo {volume} {69}},\ \bibinfo {pages} {2863--2866} (\bibinfo {year} {1992})}\BibitemShut {NoStop}%
\bibitem [{\citenamefont {Schollwöck}(2011)}]{DMRG2011}%
  \BibitemOpen
  \bibfield  {author} {\bibinfo {author} {\bibfnamefont {Ulrich}\ \bibnamefont {Schollwöck}},\ }\bibfield  {title} {\enquote {\bibinfo {title} {The density-matrix renormalization group in the age of matrix product states},}\ }\href {\doibase 10.1016/j.aop.2010.09.012} {\bibfield  {journal} {\bibinfo  {journal} {Annals of Physics}\ }\textbf {\bibinfo {volume} {326}},\ \bibinfo {pages} {96–192} (\bibinfo {year} {2011})}\BibitemShut {NoStop}%
\bibitem [{\citenamefont {Yan}\ \emph {et~al.}(2023)\citenamefont {Yan}, \citenamefont {Wang}, \citenamefont {Samajdar}, \citenamefont {Sachdev},\ and\ \citenamefont {Meng}}]{Yan2023}%
  \BibitemOpen
  \bibfield  {author} {\bibinfo {author} {\bibfnamefont {Zheng}\ \bibnamefont {Yan}}, \bibinfo {author} {\bibfnamefont {Yan-Cheng}\ \bibnamefont {Wang}}, \bibinfo {author} {\bibfnamefont {Rhine}\ \bibnamefont {Samajdar}}, \bibinfo {author} {\bibfnamefont {Subir}\ \bibnamefont {Sachdev}}, \ and\ \bibinfo {author} {\bibfnamefont {Zi~Yang}\ \bibnamefont {Meng}},\ }\bibfield  {title} {\enquote {\bibinfo {title} {Emergent glassy behavior in a kagome rydberg atom array},}\ }\href {\doibase 10.1103/PhysRevLett.130.206501} {\bibfield  {journal} {\bibinfo  {journal} {Phys. Rev. Lett.}\ }\textbf {\bibinfo {volume} {130}},\ \bibinfo {pages} {206501} (\bibinfo {year} {2023})}\BibitemShut {NoStop}%
\bibitem [{\citenamefont {Hibat-Allah}\ \emph {et~al.}(2020)\citenamefont {Hibat-Allah}, \citenamefont {Ganahl}, \citenamefont {Hayward}, \citenamefont {Melko},\ and\ \citenamefont {Carrasquilla}}]{RNNWF}%
  \BibitemOpen
  \bibfield  {author} {\bibinfo {author} {\bibfnamefont {Mohamed}\ \bibnamefont {Hibat-Allah}}, \bibinfo {author} {\bibfnamefont {Martin}\ \bibnamefont {Ganahl}}, \bibinfo {author} {\bibfnamefont {Lauren~E.}\ \bibnamefont {Hayward}}, \bibinfo {author} {\bibfnamefont {Roger~G.}\ \bibnamefont {Melko}}, \ and\ \bibinfo {author} {\bibfnamefont {Juan}\ \bibnamefont {Carrasquilla}},\ }\bibfield  {title} {\enquote {\bibinfo {title} {Recurrent neural network wave functions},}\ }\href {\doibase 10.1103/physrevresearch.2.023358} {\bibfield  {journal} {\bibinfo  {journal} {Physical Review Research}\ }\textbf {\bibinfo {volume} {2}} (\bibinfo {year} {2020}),\ 10.1103/physrevresearch.2.023358}\BibitemShut {NoStop}%
\bibitem [{\citenamefont {Roth}(2020)}]{roth2020iterative}%
  \BibitemOpen
  \bibfield  {author} {\bibinfo {author} {\bibfnamefont {Christopher}\ \bibnamefont {Roth}},\ }\href@noop {} {\enquote {\bibinfo {title} {Iterative retraining of quantum spin models using recurrent neural networks},}\ } (\bibinfo {year} {2020}),\ \Eprint {http://arxiv.org/abs/2003.06228} {arXiv:2003.06228 [physics.comp-ph]} \BibitemShut {NoStop}%
\bibitem [{\citenamefont {Moss}\ \emph {et~al.}(2023)\citenamefont {Moss}, \citenamefont {Ebadi}, \citenamefont {Wang}, \citenamefont {Semeghini}, \citenamefont {Bohrdt}, \citenamefont {Lukin},\ and\ \citenamefont {Melko}}]{moss2023enhancing}%
  \BibitemOpen
  \bibfield  {author} {\bibinfo {author} {\bibfnamefont {M.~Schuyler}\ \bibnamefont {Moss}}, \bibinfo {author} {\bibfnamefont {Sepehr}\ \bibnamefont {Ebadi}}, \bibinfo {author} {\bibfnamefont {Tout~T.}\ \bibnamefont {Wang}}, \bibinfo {author} {\bibfnamefont {Giulia}\ \bibnamefont {Semeghini}}, \bibinfo {author} {\bibfnamefont {Annabelle}\ \bibnamefont {Bohrdt}}, \bibinfo {author} {\bibfnamefont {Mikhail~D.}\ \bibnamefont {Lukin}}, \ and\ \bibinfo {author} {\bibfnamefont {Roger~G.}\ \bibnamefont {Melko}},\ }\href@noop {} {\enquote {\bibinfo {title} {Enhancing variational monte carlo using a programmable quantum simulator},}\ } (\bibinfo {year} {2023}),\ \Eprint {http://arxiv.org/abs/2308.02647} {arXiv:2308.02647 [cond-mat.quant-gas]} \BibitemShut {NoStop}%
\bibitem [{\citenamefont {Sprague}\ and\ \citenamefont {Czischek}(2023)}]{sprague2023variational}%
  \BibitemOpen
  \bibfield  {author} {\bibinfo {author} {\bibfnamefont {Kyle}\ \bibnamefont {Sprague}}\ and\ \bibinfo {author} {\bibfnamefont {Stefanie}\ \bibnamefont {Czischek}},\ }\href@noop {} {\enquote {\bibinfo {title} {Variational monte carlo with large patched transformers},}\ } (\bibinfo {year} {2023}),\ \Eprint {http://arxiv.org/abs/2306.03921} {arXiv:2306.03921 [quant-ph]} \BibitemShut {NoStop}%
\bibitem [{\citenamefont {Czischek}\ \emph {et~al.}(2022)\citenamefont {Czischek}, \citenamefont {Moss}, \citenamefont {Radzihovsky}, \citenamefont {Merali},\ and\ \citenamefont {Melko}}]{Czischek_2022}%
  \BibitemOpen
  \bibfield  {author} {\bibinfo {author} {\bibfnamefont {Stefanie}\ \bibnamefont {Czischek}}, \bibinfo {author} {\bibfnamefont {M.~Schuyler}\ \bibnamefont {Moss}}, \bibinfo {author} {\bibfnamefont {Matthew}\ \bibnamefont {Radzihovsky}}, \bibinfo {author} {\bibfnamefont {Ejaaz}\ \bibnamefont {Merali}}, \ and\ \bibinfo {author} {\bibfnamefont {Roger~G.}\ \bibnamefont {Melko}},\ }\bibfield  {title} {\enquote {\bibinfo {title} {Data-enhanced variational monte carlo simulations for rydberg atom arrays},}\ }\href {\doibase 10.1103/physrevb.105.205108} {\bibfield  {journal} {\bibinfo  {journal} {Physical Review B}\ }\textbf {\bibinfo {volume} {105}} (\bibinfo {year} {2022}),\ 10.1103/physrevb.105.205108}\BibitemShut {NoStop}%
\bibitem [{\citenamefont {Hibat-Allah}\ \emph {et~al.}(2022)\citenamefont {Hibat-Allah}, \citenamefont {Melko},\ and\ \citenamefont {Carrasquilla}}]{RNN_Symmetry_Annealing}%
  \BibitemOpen
  \bibfield  {author} {\bibinfo {author} {\bibfnamefont {Mohamed}\ \bibnamefont {Hibat-Allah}}, \bibinfo {author} {\bibfnamefont {Roger~G.}\ \bibnamefont {Melko}}, \ and\ \bibinfo {author} {\bibfnamefont {Juan}\ \bibnamefont {Carrasquilla}},\ }\href@noop {} {\enquote {\bibinfo {title} {Supplementing recurrent neural network wave functions with symmetry and annealing to improve accuracy},}\ } (\bibinfo {year} {2022}),\ \Eprint {http://arxiv.org/abs/2207.14314} {arXiv:2207.14314 [cond-mat.dis-nn]} \BibitemShut {NoStop}%
\bibitem [{\citenamefont {Bravyi}(2015)}]{bravyi2015monte}%
  \BibitemOpen
  \bibfield  {author} {\bibinfo {author} {\bibfnamefont {Sergey}\ \bibnamefont {Bravyi}},\ }\href@noop {} {\enquote {\bibinfo {title} {Monte carlo simulation of stoquastic hamiltonians},}\ } (\bibinfo {year} {2015}),\ \Eprint {http://arxiv.org/abs/1402.2295} {arXiv:1402.2295 [quant-ph]} \BibitemShut {NoStop}%
\bibitem [{\citenamefont {Lipton}\ \emph {et~al.}(2015)\citenamefont {Lipton}, \citenamefont {Berkowitz},\ and\ \citenamefont {Elkan}}]{lipton2015}%
  \BibitemOpen
  \bibfield  {author} {\bibinfo {author} {\bibfnamefont {Zachary~C.}\ \bibnamefont {Lipton}}, \bibinfo {author} {\bibfnamefont {John}\ \bibnamefont {Berkowitz}}, \ and\ \bibinfo {author} {\bibfnamefont {Charles}\ \bibnamefont {Elkan}},\ }\href@noop {} {\enquote {\bibinfo {title} {A critical review of recurrent neural networks for sequence learning},}\ } (\bibinfo {year} {2015}),\ \Eprint {http://arxiv.org/abs/1506.00019} {arXiv:1506.00019 [cs.LG]} \BibitemShut {NoStop}%
\bibitem [{\citenamefont {Hibat-Allah}\ \emph {et~al.}(2023)\citenamefont {Hibat-Allah}, \citenamefont {Melko},\ and\ \citenamefont {Carrasquilla}}]{RNN_topological_order}%
  \BibitemOpen
  \bibfield  {author} {\bibinfo {author} {\bibfnamefont {Mohamed}\ \bibnamefont {Hibat-Allah}}, \bibinfo {author} {\bibfnamefont {Roger~G.}\ \bibnamefont {Melko}}, \ and\ \bibinfo {author} {\bibfnamefont {Juan}\ \bibnamefont {Carrasquilla}},\ }\bibfield  {title} {\enquote {\bibinfo {title} {Investigating topological order using recurrent neural networks},}\ }\href {\doibase 10.1103/PhysRevB.108.075152} {\bibfield  {journal} {\bibinfo  {journal} {Phys. Rev. B}\ }\textbf {\bibinfo {volume} {108}},\ \bibinfo {pages} {075152} (\bibinfo {year} {2023})}\BibitemShut {NoStop}%
\bibitem [{\citenamefont {Casert}\ \emph {et~al.}(2021)\citenamefont {Casert}, \citenamefont {Vieijra}, \citenamefont {Whitelam},\ and\ \citenamefont {Tamblyn}}]{Vieijra2021}%
  \BibitemOpen
  \bibfield  {author} {\bibinfo {author} {\bibfnamefont {Corneel}\ \bibnamefont {Casert}}, \bibinfo {author} {\bibfnamefont {Tom}\ \bibnamefont {Vieijra}}, \bibinfo {author} {\bibfnamefont {Stephen}\ \bibnamefont {Whitelam}}, \ and\ \bibinfo {author} {\bibfnamefont {Isaac}\ \bibnamefont {Tamblyn}},\ }\bibfield  {title} {\enquote {\bibinfo {title} {Dynamical large deviations of two-dimensional kinetically constrained models using a neural-network state ansatz},}\ }\href {\doibase 10.1103/PhysRevLett.127.120602} {\bibfield  {journal} {\bibinfo  {journal} {Phys. Rev. Lett.}\ }\textbf {\bibinfo {volume} {127}},\ \bibinfo {pages} {120602} (\bibinfo {year} {2021})}\BibitemShut {NoStop}%
\bibitem [{\citenamefont {Luo}\ \emph {et~al.}(2023)\citenamefont {Luo}, \citenamefont {Chen}, \citenamefont {Hu}, \citenamefont {Zhao}, \citenamefont {Hur},\ and\ \citenamefont {Clark}}]{luo2021gauge}%
  \BibitemOpen
  \bibfield  {author} {\bibinfo {author} {\bibfnamefont {Di}~\bibnamefont {Luo}}, \bibinfo {author} {\bibfnamefont {Zhuo}\ \bibnamefont {Chen}}, \bibinfo {author} {\bibfnamefont {Kaiwen}\ \bibnamefont {Hu}}, \bibinfo {author} {\bibfnamefont {Zhizhen}\ \bibnamefont {Zhao}}, \bibinfo {author} {\bibfnamefont {Vera~Mikyoung}\ \bibnamefont {Hur}}, \ and\ \bibinfo {author} {\bibfnamefont {Bryan~K.}\ \bibnamefont {Clark}},\ }\bibfield  {title} {\enquote {\bibinfo {title} {Gauge-invariant and anyonic-symmetric autoregressive neural network for quantum lattice models},}\ }\href {\doibase 10.1103/PhysRevResearch.5.013216} {\bibfield  {journal} {\bibinfo  {journal} {Phys. Rev. Res.}\ }\textbf {\bibinfo {volume} {5}},\ \bibinfo {pages} {013216} (\bibinfo {year} {2023})}\BibitemShut {NoStop}%
\bibitem [{\citenamefont {Becca}\ and\ \citenamefont {Sorella}(2017)}]{becca_sorella_2017}%
  \BibitemOpen
  \bibfield  {author} {\bibinfo {author} {\bibfnamefont {Federico}\ \bibnamefont {Becca}}\ and\ \bibinfo {author} {\bibfnamefont {Sandro}\ \bibnamefont {Sorella}},\ }\href {\doibase 10.1017/9781316417041} {\emph {\bibinfo {title} {Quantum Monte Carlo Approaches for Correlated Systems}}}\ (\bibinfo  {publisher} {Cambridge University Press},\ \bibinfo {year} {2017})\BibitemShut {NoStop}%
\bibitem [{\citenamefont {Hibat-Allah}\ \emph {et~al.}(2021)\citenamefont {Hibat-Allah}, \citenamefont {Inack}, \citenamefont {Wiersema}, \citenamefont {Melko},\ and\ \citenamefont {Carrasquilla}}]{VNA2021}%
  \BibitemOpen
  \bibfield  {author} {\bibinfo {author} {\bibfnamefont {Mohamed}\ \bibnamefont {Hibat-Allah}}, \bibinfo {author} {\bibfnamefont {Estelle~M.}\ \bibnamefont {Inack}}, \bibinfo {author} {\bibfnamefont {Roeland}\ \bibnamefont {Wiersema}}, \bibinfo {author} {\bibfnamefont {Roger~G.}\ \bibnamefont {Melko}}, \ and\ \bibinfo {author} {\bibfnamefont {Juan}\ \bibnamefont {Carrasquilla}},\ }\bibfield  {title} {\enquote {\bibinfo {title} {Variational neural annealing},}\ }\href {\doibase 10.1038/s42256-021-00401-3} {\bibfield  {journal} {\bibinfo  {journal} {Nature Machine Intelligence}\ } (\bibinfo {year} {2021}),\ 10.1038/s42256-021-00401-3}\BibitemShut {NoStop}%
\bibitem [{\citenamefont {Roth}\ \emph {et~al.}(2022)\citenamefont {Roth}, \citenamefont {Szabó},\ and\ \citenamefont {MacDonald}}]{Roth2022}%
  \BibitemOpen
  \bibfield  {author} {\bibinfo {author} {\bibfnamefont {Christopher}\ \bibnamefont {Roth}}, \bibinfo {author} {\bibfnamefont {Attila}\ \bibnamefont {Szabó}}, \ and\ \bibinfo {author} {\bibfnamefont {Allan}\ \bibnamefont {MacDonald}},\ }\href {\doibase 10.48550/ARXIV.2211.07749} {\enquote {\bibinfo {title} {High-accuracy variational monte carlo for frustrated magnets with deep neural networks},}\ } (\bibinfo {year} {2022})\BibitemShut {NoStop}%
\bibitem [{\citenamefont {Khandoker}\ \emph {et~al.}(2023)\citenamefont {Khandoker}, \citenamefont {Abedin},\ and\ \citenamefont {Hibat-Allah}}]{RNNOpt2023}%
  \BibitemOpen
  \bibfield  {author} {\bibinfo {author} {\bibfnamefont {Shoummo~Ahsan}\ \bibnamefont {Khandoker}}, \bibinfo {author} {\bibfnamefont {Jawaril~Munshad}\ \bibnamefont {Abedin}}, \ and\ \bibinfo {author} {\bibfnamefont {Mohamed}\ \bibnamefont {Hibat-Allah}},\ }\bibfield  {title} {\enquote {\bibinfo {title} {Supplementing recurrent neural networks with annealing to solve combinatorial optimization problems},}\ }\href {\doibase 10.1088/2632-2153/acb895} {\bibfield  {journal} {\bibinfo  {journal} {Machine Learning: Science and Technology}\ }\textbf {\bibinfo {volume} {4}},\ \bibinfo {pages} {015026} (\bibinfo {year} {2023})}\BibitemShut {NoStop}%
\bibitem [{\citenamefont {Hamma}\ \emph {et~al.}(2005{\natexlab{a}})\citenamefont {Hamma}, \citenamefont {Ionicioiu},\ and\ \citenamefont {Zanardi}}]{Hamma2004}%
  \BibitemOpen
  \bibfield  {author} {\bibinfo {author} {\bibfnamefont {Alioscia}\ \bibnamefont {Hamma}}, \bibinfo {author} {\bibfnamefont {Radu}\ \bibnamefont {Ionicioiu}}, \ and\ \bibinfo {author} {\bibfnamefont {Paolo}\ \bibnamefont {Zanardi}},\ }\bibfield  {title} {\enquote {\bibinfo {title} {Bipartite entanglement and entropic boundary law in lattice spin systems},}\ }\href {\doibase 10.1103/physreva.71.022315} {\bibfield  {journal} {\bibinfo  {journal} {Physical Review A}\ }\textbf {\bibinfo {volume} {71}} (\bibinfo {year} {2005}{\natexlab{a}}),\ 10.1103/physreva.71.022315}\BibitemShut {NoStop}%
\bibitem [{\citenamefont {Hamma}\ \emph {et~al.}(2005{\natexlab{b}})\citenamefont {Hamma}, \citenamefont {Ionicioiu},\ and\ \citenamefont {Zanardi}}]{Hamma2005}%
  \BibitemOpen
  \bibfield  {author} {\bibinfo {author} {\bibfnamefont {Alioscia}\ \bibnamefont {Hamma}}, \bibinfo {author} {\bibfnamefont {Radu}\ \bibnamefont {Ionicioiu}}, \ and\ \bibinfo {author} {\bibfnamefont {Paolo}\ \bibnamefont {Zanardi}},\ }\bibfield  {title} {\enquote {\bibinfo {title} {Ground state entanglement and geometric entropy in the kitaev model},}\ }\href {\doibase 10.1016/j.physleta.2005.01.060} {\bibfield  {journal} {\bibinfo  {journal} {Physics Letters A}\ }\textbf {\bibinfo {volume} {337}},\ \bibinfo {pages} {22–28} (\bibinfo {year} {2005}{\natexlab{b}})}\BibitemShut {NoStop}%
\bibitem [{\citenamefont {Levin}\ and\ \citenamefont {Wen}(2006)}]{LevinWen2006}%
  \BibitemOpen
  \bibfield  {author} {\bibinfo {author} {\bibfnamefont {Michael}\ \bibnamefont {Levin}}\ and\ \bibinfo {author} {\bibfnamefont {Xiao-Gang}\ \bibnamefont {Wen}},\ }\bibfield  {title} {\enquote {\bibinfo {title} {Detecting topological order in a ground state wave function},}\ }\href {\doibase 10.1103/physrevlett.96.110405} {\bibfield  {journal} {\bibinfo  {journal} {Physical Review Letters}\ }\textbf {\bibinfo {volume} {96}} (\bibinfo {year} {2006}),\ 10.1103/physrevlett.96.110405}\BibitemShut {NoStop}%
\bibitem [{\citenamefont {Kitaev}\ and\ \citenamefont {Preskill}(2006)}]{KitaevPreskill2006}%
  \BibitemOpen
  \bibfield  {author} {\bibinfo {author} {\bibfnamefont {Alexei}\ \bibnamefont {Kitaev}}\ and\ \bibinfo {author} {\bibfnamefont {John}\ \bibnamefont {Preskill}},\ }\bibfield  {title} {\enquote {\bibinfo {title} {Topological entanglement entropy},}\ }\href {\doibase 10.1103/PhysRevLett.96.110404} {\bibfield  {journal} {\bibinfo  {journal} {Phys. Rev. Lett.}\ }\textbf {\bibinfo {volume} {96}},\ \bibinfo {pages} {110404} (\bibinfo {year} {2006})}\BibitemShut {NoStop}%
\bibitem [{\citenamefont {Flammia}\ \emph {et~al.}(2009)\citenamefont {Flammia}, \citenamefont {Hamma}, \citenamefont {Hughes},\ and\ \citenamefont {Wen}}]{Hamma2009}%
  \BibitemOpen
  \bibfield  {author} {\bibinfo {author} {\bibfnamefont {Steven~T.}\ \bibnamefont {Flammia}}, \bibinfo {author} {\bibfnamefont {Alioscia}\ \bibnamefont {Hamma}}, \bibinfo {author} {\bibfnamefont {Taylor~L.}\ \bibnamefont {Hughes}}, \ and\ \bibinfo {author} {\bibfnamefont {Xiao-Gang}\ \bibnamefont {Wen}},\ }\bibfield  {title} {\enquote {\bibinfo {title} {Topological entanglement rényi entropy and reduced density matrix structure},}\ }\href {\doibase 10.1103/physrevlett.103.261601} {\bibfield  {journal} {\bibinfo  {journal} {Physical Review Letters}\ }\textbf {\bibinfo {volume} {103}} (\bibinfo {year} {2009}),\ 10.1103/physrevlett.103.261601}\BibitemShut {NoStop}%
\bibitem [{\citenamefont {Isakov}\ \emph {et~al.}(2011)\citenamefont {Isakov}, \citenamefont {Hastings},\ and\ \citenamefont {Melko}}]{TEE2011}%
  \BibitemOpen
  \bibfield  {author} {\bibinfo {author} {\bibfnamefont {Sergei~V.}\ \bibnamefont {Isakov}}, \bibinfo {author} {\bibfnamefont {Matthew~B.}\ \bibnamefont {Hastings}}, \ and\ \bibinfo {author} {\bibfnamefont {Roger~G.}\ \bibnamefont {Melko}},\ }\bibfield  {title} {\enquote {\bibinfo {title} {Topological entanglement entropy of a bose–hubbard spin liquid},}\ }\href {\doibase 10.1038/nphys2036} {\bibfield  {journal} {\bibinfo  {journal} {Nature Physics}\ }\textbf {\bibinfo {volume} {7}},\ \bibinfo {pages} {772–775} (\bibinfo {year} {2011})}\BibitemShut {NoStop}%
\bibitem [{\citenamefont {Wildeboer}\ \emph {et~al.}(2017)\citenamefont {Wildeboer}, \citenamefont {Seidel},\ and\ \citenamefont {Melko}}]{TEE2017}%
  \BibitemOpen
  \bibfield  {author} {\bibinfo {author} {\bibfnamefont {Julia}\ \bibnamefont {Wildeboer}}, \bibinfo {author} {\bibfnamefont {Alexander}\ \bibnamefont {Seidel}}, \ and\ \bibinfo {author} {\bibfnamefont {Roger~G.}\ \bibnamefont {Melko}},\ }\bibfield  {title} {\enquote {\bibinfo {title} {Entanglement entropy and topological order in resonating valence-bond quantum spin liquids},}\ }\href {\doibase 10.1103/physrevb.95.100402} {\bibfield  {journal} {\bibinfo  {journal} {Physical Review B}\ }\textbf {\bibinfo {volume} {95}} (\bibinfo {year} {2017}),\ 10.1103/physrevb.95.100402}\BibitemShut {NoStop}%
\bibitem [{\citenamefont {Hastings}\ \emph {et~al.}(2010)\citenamefont {Hastings}, \citenamefont {González}, \citenamefont {Kallin},\ and\ \citenamefont {Melko}}]{EE2010}%
  \BibitemOpen
  \bibfield  {author} {\bibinfo {author} {\bibfnamefont {Matthew~B.}\ \bibnamefont {Hastings}}, \bibinfo {author} {\bibfnamefont {Iván}\ \bibnamefont {González}}, \bibinfo {author} {\bibfnamefont {Ann~B.}\ \bibnamefont {Kallin}}, \ and\ \bibinfo {author} {\bibfnamefont {Roger~G.}\ \bibnamefont {Melko}},\ }\bibfield  {title} {\enquote {\bibinfo {title} {Measuring renyi entanglement entropy in quantum monte carlo simulations},}\ }\href {\doibase 10.1103/physrevlett.104.157201} {\bibfield  {journal} {\bibinfo  {journal} {Physical Review Letters}\ }\textbf {\bibinfo {volume} {104}} (\bibinfo {year} {2010}),\ 10.1103/physrevlett.104.157201}\BibitemShut {NoStop}%
\bibitem [{\citenamefont {Wang}\ and\ \citenamefont {Davis}(2020)}]{EE2020}%
  \BibitemOpen
  \bibfield  {author} {\bibinfo {author} {\bibfnamefont {Zhaoyou}\ \bibnamefont {Wang}}\ and\ \bibinfo {author} {\bibfnamefont {Emily~J.}\ \bibnamefont {Davis}},\ }\bibfield  {title} {\enquote {\bibinfo {title} {Calculating rényi entropies with neural autoregressive quantum states},}\ }\href {\doibase 10.1103/physreva.102.062413} {\bibfield  {journal} {\bibinfo  {journal} {Physical Review A}\ }\textbf {\bibinfo {volume} {102}} (\bibinfo {year} {2020}),\ 10.1103/physreva.102.062413}\BibitemShut {NoStop}%
\bibitem [{\citenamefont {Furukawa}\ and\ \citenamefont {Misguich}(2007)}]{Furukawa2007}%
  \BibitemOpen
  \bibfield  {author} {\bibinfo {author} {\bibfnamefont {Shunsuke}\ \bibnamefont {Furukawa}}\ and\ \bibinfo {author} {\bibfnamefont {Grégoire}\ \bibnamefont {Misguich}},\ }\bibfield  {title} {\enquote {\bibinfo {title} {Topological entanglement entropy in the quantum dimer model on the triangular lattice},}\ }\href {\doibase 10.1103/physrevb.75.214407} {\bibfield  {journal} {\bibinfo  {journal} {Physical Review B}\ }\textbf {\bibinfo {volume} {75}} (\bibinfo {year} {2007}),\ 10.1103/physrevb.75.214407}\BibitemShut {NoStop}%
\bibitem [{\citenamefont {Stoudenmire}\ and\ \citenamefont {White}(2012)}]{Stoudenmire2012}%
  \BibitemOpen
  \bibfield  {author} {\bibinfo {author} {\bibfnamefont {E.M.}\ \bibnamefont {Stoudenmire}}\ and\ \bibinfo {author} {\bibfnamefont {Steven~R.}\ \bibnamefont {White}},\ }\bibfield  {title} {\enquote {\bibinfo {title} {Studying two-dimensional systems with the density matrix renormalization group},}\ }\href {\doibase 10.1146/annurev-conmatphys-020911-125018} {\bibfield  {journal} {\bibinfo  {journal} {Annual Review of Condensed Matter Physics}\ }\textbf {\bibinfo {volume} {3}},\ \bibinfo {pages} {111–128} (\bibinfo {year} {2012})}\BibitemShut {NoStop}%
\bibitem [{\citenamefont {Gong}\ \emph {et~al.}(2014)\citenamefont {Gong}, \citenamefont {Zhu}, \citenamefont {Sheng}, \citenamefont {Motrunich},\ and\ \citenamefont {Fisher}}]{Cylinders2014}%
  \BibitemOpen
  \bibfield  {author} {\bibinfo {author} {\bibfnamefont {Shou-Shu}\ \bibnamefont {Gong}}, \bibinfo {author} {\bibfnamefont {Wei}\ \bibnamefont {Zhu}}, \bibinfo {author} {\bibfnamefont {D.~N.}\ \bibnamefont {Sheng}}, \bibinfo {author} {\bibfnamefont {Olexei~I.}\ \bibnamefont {Motrunich}}, \ and\ \bibinfo {author} {\bibfnamefont {Matthew P.~A.}\ \bibnamefont {Fisher}},\ }\bibfield  {title} {\enquote {\bibinfo {title} {Plaquette ordered phase and quantum phase diagram in the spin-$\frac{1}{2}$ ${J}_{1}\text{\ensuremath{-}}{J}_{2}$ square heisenberg model},}\ }\href {\doibase 10.1103/PhysRevLett.113.027201} {\bibfield  {journal} {\bibinfo  {journal} {Phys. Rev. Lett.}\ }\textbf {\bibinfo {volume} {113}},\ \bibinfo {pages} {027201} (\bibinfo {year} {2014})}\BibitemShut {NoStop}%
\bibitem [{\citenamefont {Nikoli\ifmmode~\acute{c}\else \'{c}\fi{}}\ and\ \citenamefont {Senthil}(2005)}]{Nikolic2005}%
  \BibitemOpen
  \bibfield  {author} {\bibinfo {author} {\bibfnamefont {P.}~\bibnamefont {Nikoli\ifmmode~\acute{c}\else \'{c}\fi{}}}\ and\ \bibinfo {author} {\bibfnamefont {T.}~\bibnamefont {Senthil}},\ }\bibfield  {title} {\enquote {\bibinfo {title} {Theory of the kagome lattice ising antiferromagnet in weak transverse fields},}\ }\href {\doibase 10.1103/PhysRevB.71.024401} {\bibfield  {journal} {\bibinfo  {journal} {Phys. Rev. B}\ }\textbf {\bibinfo {volume} {71}},\ \bibinfo {pages} {024401} (\bibinfo {year} {2005})}\BibitemShut {NoStop}%
\bibitem [{\citenamefont {Moessner}\ and\ \citenamefont {Sondhi}(2001)}]{Moessner2001}%
  \BibitemOpen
  \bibfield  {author} {\bibinfo {author} {\bibfnamefont {R.}~\bibnamefont {Moessner}}\ and\ \bibinfo {author} {\bibfnamefont {S.~L.}\ \bibnamefont {Sondhi}},\ }\bibfield  {title} {\enquote {\bibinfo {title} {Ising models of quantum frustration},}\ }\href {\doibase 10.1103/PhysRevB.63.224401} {\bibfield  {journal} {\bibinfo  {journal} {Phys. Rev. B}\ }\textbf {\bibinfo {volume} {63}},\ \bibinfo {pages} {224401} (\bibinfo {year} {2001})}\BibitemShut {NoStop}%
\bibitem [{\citenamefont {Moessner}\ \emph {et~al.}(2000)\citenamefont {Moessner}, \citenamefont {Sondhi},\ and\ \citenamefont {Chandra}}]{Moessner_2000}%
  \BibitemOpen
  \bibfield  {author} {\bibinfo {author} {\bibfnamefont {R.}~\bibnamefont {Moessner}}, \bibinfo {author} {\bibfnamefont {S.~L.}\ \bibnamefont {Sondhi}}, \ and\ \bibinfo {author} {\bibfnamefont {P.}~\bibnamefont {Chandra}},\ }\bibfield  {title} {\enquote {\bibinfo {title} {Two-dimensional periodic frustrated ising models in a transverse field},}\ }\href {\doibase 10.1103/physrevlett.84.4457} {\bibfield  {journal} {\bibinfo  {journal} {Physical Review Letters}\ }\textbf {\bibinfo {volume} {84}},\ \bibinfo {pages} {4457–4460} (\bibinfo {year} {2000})}\BibitemShut {NoStop}%
\bibitem [{\citenamefont {Edwards}\ and\ \citenamefont {Anderson}(1975)}]{EA_1975}%
  \BibitemOpen
  \bibfield  {author} {\bibinfo {author} {\bibfnamefont {S~F}\ \bibnamefont {Edwards}}\ and\ \bibinfo {author} {\bibfnamefont {P~W}\ \bibnamefont {Anderson}},\ }\bibfield  {title} {\enquote {\bibinfo {title} {Theory of spin glasses},}\ }\href {\doibase 10.1088/0305-4608/5/5/017} {\bibfield  {journal} {\bibinfo  {journal} {Journal of Physics F: Metal Physics}\ }\textbf {\bibinfo {volume} {5}},\ \bibinfo {pages} {965} (\bibinfo {year} {1975})}\BibitemShut {NoStop}%
\bibitem [{\citenamefont {Richards}(1984)}]{Richards84}%
  \BibitemOpen
  \bibfield  {author} {\bibinfo {author} {\bibfnamefont {Peter~M.}\ \bibnamefont {Richards}},\ }\bibfield  {title} {\enquote {\bibinfo {title} {Spin-glass order parameter of the random-field ising model},}\ }\href {\doibase 10.1103/PhysRevB.30.2955} {\bibfield  {journal} {\bibinfo  {journal} {Phys. Rev. B}\ }\textbf {\bibinfo {volume} {30}},\ \bibinfo {pages} {2955--2957} (\bibinfo {year} {1984})}\BibitemShut {NoStop}%
\bibitem [{\citenamefont {Castellani}\ and\ \citenamefont {Cavagna}(2005)}]{Castellani_2005}%
  \BibitemOpen
  \bibfield  {author} {\bibinfo {author} {\bibfnamefont {Tommaso}\ \bibnamefont {Castellani}}\ and\ \bibinfo {author} {\bibfnamefont {Andrea}\ \bibnamefont {Cavagna}},\ }\bibfield  {title} {\enquote {\bibinfo {title} {Spin-glass theory for pedestrians},}\ }\href {\doibase 10.1088/1742-5468/2005/05/p05012} {\bibfield  {journal} {\bibinfo  {journal} {Journal of Statistical Mechanics: Theory and Experiment}\ }\textbf {\bibinfo {volume} {2005}},\ \bibinfo {pages} {P05012} (\bibinfo {year} {2005})}\BibitemShut {NoStop}%
\bibitem [{\citenamefont {Merali}\ \emph {et~al.}(2023)\citenamefont {Merali}, \citenamefont {Vlugt},\ and\ \citenamefont {Melko}}]{merali2023stochastic}%
  \BibitemOpen
  \bibfield  {author} {\bibinfo {author} {\bibfnamefont {Ejaaz}\ \bibnamefont {Merali}}, \bibinfo {author} {\bibfnamefont {Isaac J. S.~De}\ \bibnamefont {Vlugt}}, \ and\ \bibinfo {author} {\bibfnamefont {Roger~G.}\ \bibnamefont {Melko}},\ }\href@noop {} {\enquote {\bibinfo {title} {Stochastic series expansion quantum monte carlo for rydberg arrays},}\ } (\bibinfo {year} {2023}),\ \Eprint {http://arxiv.org/abs/2107.00766} {arXiv:2107.00766 [cond-mat.str-el]} \BibitemShut {NoStop}%
\bibitem [{\citenamefont {Sandvik}\ and\ \citenamefont {Kurkij\"arvi}(1991)}]{PhysRevB.43.5950}%
  \BibitemOpen
  \bibfield  {author} {\bibinfo {author} {\bibfnamefont {Anders~W.}\ \bibnamefont {Sandvik}}\ and\ \bibinfo {author} {\bibfnamefont {Juhani}\ \bibnamefont {Kurkij\"arvi}},\ }\bibfield  {title} {\enquote {\bibinfo {title} {Quantum monte carlo simulation method for spin systems},}\ }\href {\doibase 10.1103/PhysRevB.43.5950} {\bibfield  {journal} {\bibinfo  {journal} {Phys. Rev. B}\ }\textbf {\bibinfo {volume} {43}},\ \bibinfo {pages} {5950--5961} (\bibinfo {year} {1991})}\BibitemShut {NoStop}%
\bibitem [{\citenamefont {Sandvik}(1999)}]{Sandvik_1999}%
  \BibitemOpen
  \bibfield  {author} {\bibinfo {author} {\bibfnamefont {Anders~W.}\ \bibnamefont {Sandvik}},\ }\bibfield  {title} {\enquote {\bibinfo {title} {Stochastic series expansion method with operator-loop update},}\ }\href {\doibase 10.1103/physrevb.59.r14157} {\bibfield  {journal} {\bibinfo  {journal} {Physical Review B}\ }\textbf {\bibinfo {volume} {59}},\ \bibinfo {pages} {R14157–R14160} (\bibinfo {year} {1999})}\BibitemShut {NoStop}%
\bibitem [{\citenamefont {Yan}\ \emph {et~al.}(2022)\citenamefont {Yan}, \citenamefont {Samajdar}, \citenamefont {Wang}, \citenamefont {Sachdev},\ and\ \citenamefont {Meng}}]{Yan2022}%
  \BibitemOpen
  \bibfield  {author} {\bibinfo {author} {\bibfnamefont {Zheng}\ \bibnamefont {Yan}}, \bibinfo {author} {\bibfnamefont {Rhine}\ \bibnamefont {Samajdar}}, \bibinfo {author} {\bibfnamefont {Yan-Cheng}\ \bibnamefont {Wang}}, \bibinfo {author} {\bibfnamefont {Subir}\ \bibnamefont {Sachdev}}, \ and\ \bibinfo {author} {\bibfnamefont {Zi~Yang}\ \bibnamefont {Meng}},\ }\bibfield  {title} {\enquote {\bibinfo {title} {Triangular lattice quantum dimer model with variable dimer density},}\ }\href {\doibase 10.1038/s41467-022-33431-5} {\bibfield  {journal} {\bibinfo  {journal} {Nature Communications}\ }\textbf {\bibinfo {volume} {13}},\ \bibinfo {pages} {5799} (\bibinfo {year} {2022})}\BibitemShut {NoStop}%
\bibitem [{\citenamefont {Wu}\ \emph {et~al.}(2024)\citenamefont {Wu}, \citenamefont {Rossi}, \citenamefont {Vicentini}, \citenamefont {Astrakhantsev}, \citenamefont {Becca}, \citenamefont {Cao}, \citenamefont {Carrasquilla}, \citenamefont {Ferrari}, \citenamefont {Georges}, \citenamefont {Hibat-Allah}, \citenamefont {Imada}, \citenamefont {Läuchli}, \citenamefont {Mazzola}, \citenamefont {Mezzacapo}, \citenamefont {Millis}, \citenamefont {Moreno}, \citenamefont {Neupert}, \citenamefont {Nomura}, \citenamefont {Nys}, \citenamefont {Parcollet}, \citenamefont {Pohle}, \citenamefont {Romero}, \citenamefont {Schmid}, \citenamefont {Silvester}, \citenamefont {Sorella}, \citenamefont {Tocchio}, \citenamefont {Wang}, \citenamefont {White}, \citenamefont {Wietek}, \citenamefont {Yang}, \citenamefont {Yang}, \citenamefont {Zhang},\ and\ \citenamefont {Carleo}}]{v_score}%
  \BibitemOpen
  \bibfield  {author} {\bibinfo {author} {\bibfnamefont {Dian}\ \bibnamefont {Wu}}, \bibinfo {author} {\bibfnamefont {Riccardo}\ \bibnamefont {Rossi}}, \bibinfo {author} {\bibfnamefont {Filippo}\ \bibnamefont {Vicentini}}, \bibinfo {author} {\bibfnamefont {Nikita}\ \bibnamefont {Astrakhantsev}}, \bibinfo {author} {\bibfnamefont {Federico}\ \bibnamefont {Becca}}, \bibinfo {author} {\bibfnamefont {Xiaodong}\ \bibnamefont {Cao}}, \bibinfo {author} {\bibfnamefont {Juan}\ \bibnamefont {Carrasquilla}}, \bibinfo {author} {\bibfnamefont {Francesco}\ \bibnamefont {Ferrari}}, \bibinfo {author} {\bibfnamefont {Antoine}\ \bibnamefont {Georges}}, \bibinfo {author} {\bibfnamefont {Mohamed}\ \bibnamefont {Hibat-Allah}}, \bibinfo {author} {\bibfnamefont {Masatoshi}\ \bibnamefont {Imada}}, \bibinfo {author} {\bibfnamefont {Andreas~M.}\ \bibnamefont {Läuchli}}, \bibinfo {author} {\bibfnamefont {Guglielmo}\ \bibnamefont {Mazzola}}, \bibinfo {author} {\bibfnamefont {Antonio}\ \bibnamefont {Mezzacapo}}, \bibinfo {author}
  {\bibfnamefont {Andrew}\ \bibnamefont {Millis}}, \bibinfo {author} {\bibfnamefont {Javier~Robledo}\ \bibnamefont {Moreno}}, \bibinfo {author} {\bibfnamefont {Titus}\ \bibnamefont {Neupert}}, \bibinfo {author} {\bibfnamefont {Yusuke}\ \bibnamefont {Nomura}}, \bibinfo {author} {\bibfnamefont {Jannes}\ \bibnamefont {Nys}}, \bibinfo {author} {\bibfnamefont {Olivier}\ \bibnamefont {Parcollet}}, \bibinfo {author} {\bibfnamefont {Rico}\ \bibnamefont {Pohle}}, \bibinfo {author} {\bibfnamefont {Imelda}\ \bibnamefont {Romero}}, \bibinfo {author} {\bibfnamefont {Michael}\ \bibnamefont {Schmid}}, \bibinfo {author} {\bibfnamefont {J.~Maxwell}\ \bibnamefont {Silvester}}, \bibinfo {author} {\bibfnamefont {Sandro}\ \bibnamefont {Sorella}}, \bibinfo {author} {\bibfnamefont {Luca~F.}\ \bibnamefont {Tocchio}}, \bibinfo {author} {\bibfnamefont {Lei}\ \bibnamefont {Wang}}, \bibinfo {author} {\bibfnamefont {Steven~R.}\ \bibnamefont {White}}, \bibinfo {author} {\bibfnamefont {Alexander}\ \bibnamefont {Wietek}}, \bibinfo {author}
  {\bibfnamefont {Qi}~\bibnamefont {Yang}}, \bibinfo {author} {\bibfnamefont {Yiqi}\ \bibnamefont {Yang}}, \bibinfo {author} {\bibfnamefont {Shiwei}\ \bibnamefont {Zhang}}, \ and\ \bibinfo {author} {\bibfnamefont {Giuseppe}\ \bibnamefont {Carleo}},\ }\bibfield  {title} {\enquote {\bibinfo {title} {Variational benchmarks for quantum many-body problems},}\ }\href {\doibase 10.1126/science.adg9774} {\bibfield  {journal} {\bibinfo  {journal} {Science}\ }\textbf {\bibinfo {volume} {386}},\ \bibinfo {pages} {296--301} (\bibinfo {year} {2024})},\ \Eprint {http://arxiv.org/abs/https://www.science.org/doi/pdf/10.1126/science.adg9774} {https://www.science.org/doi/pdf/10.1126/science.adg9774} \BibitemShut {NoStop}%
\bibitem [{\citenamefont {Moss}\ \emph {et~al.}(2025{\natexlab{a}})\citenamefont {Moss}, \citenamefont {Wiersema}, \citenamefont {Hibat-Allah}, \citenamefont {Carrasquilla},\ and\ \citenamefont {Melko}}]{moss2025leveragingrecurrenceneuralnetwork}%
  \BibitemOpen
  \bibfield  {author} {\bibinfo {author} {\bibfnamefont {M.~Schuyler}\ \bibnamefont {Moss}}, \bibinfo {author} {\bibfnamefont {Roeland}\ \bibnamefont {Wiersema}}, \bibinfo {author} {\bibfnamefont {Mohamed}\ \bibnamefont {Hibat-Allah}}, \bibinfo {author} {\bibfnamefont {Juan}\ \bibnamefont {Carrasquilla}}, \ and\ \bibinfo {author} {\bibfnamefont {Roger~G.}\ \bibnamefont {Melko}},\ }\href {https://arxiv.org/abs/2502.17144} {\enquote {\bibinfo {title} {Leveraging recurrence in neural network wavefunctions for large-scale simulations of heisenberg antiferromagnets: the square lattice},}\ } (\bibinfo {year} {2025}{\natexlab{a}}),\ \Eprint {http://arxiv.org/abs/2502.17144} {arXiv:2502.17144 [cond-mat.str-el]} \BibitemShut {NoStop}%
\bibitem [{\citenamefont {Moss}\ \emph {et~al.}(2025{\natexlab{b}})\citenamefont {Moss}, \citenamefont {Wiersema}, \citenamefont {Hibat-Allah}, \citenamefont {Carrasquilla},\ and\ \citenamefont {Melko}}]{moss2025_triangular}%
  \BibitemOpen
  \bibfield  {author} {\bibinfo {author} {\bibfnamefont {M.~Schuyler}\ \bibnamefont {Moss}}, \bibinfo {author} {\bibfnamefont {Roeland}\ \bibnamefont {Wiersema}}, \bibinfo {author} {\bibfnamefont {Mohamed}\ \bibnamefont {Hibat-Allah}}, \bibinfo {author} {\bibfnamefont {Juan}\ \bibnamefont {Carrasquilla}}, \ and\ \bibinfo {author} {\bibfnamefont {Roger~G.}\ \bibnamefont {Melko}},\ }\href {https://arxiv.org/abs/2505.20406} {\enquote {\bibinfo {title} {Leveraging recurrence in neural network wavefunctions for large-scale simulations of heisenberg antiferromagnets: the triangular lattice},}\ } (\bibinfo {year} {2025}{\natexlab{b}}),\ \Eprint {http://arxiv.org/abs/2505.20406} {arXiv:2505.20406 [cond-mat.str-el]} \BibitemShut {NoStop}%
\bibitem [{\citenamefont {Chen}\ and\ \citenamefont {Heyl}(2024)}]{Chen_2024}%
  \BibitemOpen
  \bibfield  {author} {\bibinfo {author} {\bibfnamefont {Ao}~\bibnamefont {Chen}}\ and\ \bibinfo {author} {\bibfnamefont {Markus}\ \bibnamefont {Heyl}},\ }\bibfield  {title} {\enquote {\bibinfo {title} {Empowering deep neural quantum states through efficient optimization},}\ }\href {\doibase 10.1038/s41567-024-02566-1} {\bibfield  {journal} {\bibinfo  {journal} {Nature Physics}\ }\textbf {\bibinfo {volume} {20}},\ \bibinfo {pages} {1476–1481} (\bibinfo {year} {2024})}\BibitemShut {NoStop}%
\bibitem [{\citenamefont {Rende}\ \emph {et~al.}(2024)\citenamefont {Rende}, \citenamefont {Viteritti}, \citenamefont {Bardone}, \citenamefont {Becca},\ and\ \citenamefont {Goldt}}]{Rende_2024}%
  \BibitemOpen
  \bibfield  {author} {\bibinfo {author} {\bibfnamefont {Riccardo}\ \bibnamefont {Rende}}, \bibinfo {author} {\bibfnamefont {Luciano~Loris}\ \bibnamefont {Viteritti}}, \bibinfo {author} {\bibfnamefont {Lorenzo}\ \bibnamefont {Bardone}}, \bibinfo {author} {\bibfnamefont {Federico}\ \bibnamefont {Becca}}, \ and\ \bibinfo {author} {\bibfnamefont {Sebastian}\ \bibnamefont {Goldt}},\ }\bibfield  {title} {\enquote {\bibinfo {title} {A simple linear algebra identity to optimize large-scale neural network quantum states},}\ }\href {\doibase 10.1038/s42005-024-01732-4} {\bibfield  {journal} {\bibinfo  {journal} {Communications Physics}\ }\textbf {\bibinfo {volume} {7}} (\bibinfo {year} {2024}),\ 10.1038/s42005-024-01732-4}\BibitemShut {NoStop}%
\bibitem [{\citenamefont {Donatella}\ \emph {et~al.}(2023)\citenamefont {Donatella}, \citenamefont {Denis}, \citenamefont {Le~Boit\'e},\ and\ \citenamefont {Ciuti}}]{donatella_autoregressive}%
  \BibitemOpen
  \bibfield  {author} {\bibinfo {author} {\bibfnamefont {Kaelan}\ \bibnamefont {Donatella}}, \bibinfo {author} {\bibfnamefont {Zakari}\ \bibnamefont {Denis}}, \bibinfo {author} {\bibfnamefont {Alexandre}\ \bibnamefont {Le~Boit\'e}}, \ and\ \bibinfo {author} {\bibfnamefont {Cristiano}\ \bibnamefont {Ciuti}},\ }\bibfield  {title} {\enquote {\bibinfo {title} {Dynamics with autoregressive neural quantum states: Application to critical quench dynamics},}\ }\href {\doibase 10.1103/PhysRevA.108.022210} {\bibfield  {journal} {\bibinfo  {journal} {Phys. Rev. A}\ }\textbf {\bibinfo {volume} {108}},\ \bibinfo {pages} {022210} (\bibinfo {year} {2023})}\BibitemShut {NoStop}%
\bibitem [{\citenamefont {Lange}\ \emph {et~al.}(2024)\citenamefont {Lange}, \citenamefont {D{\"o}schl}, \citenamefont {Carrasquilla},\ and\ \citenamefont {Bohrdt}}]{lange2024neural}%
  \BibitemOpen
  \bibfield  {author} {\bibinfo {author} {\bibfnamefont {Hannah}\ \bibnamefont {Lange}}, \bibinfo {author} {\bibfnamefont {Fabian}\ \bibnamefont {D{\"o}schl}}, \bibinfo {author} {\bibfnamefont {Juan}\ \bibnamefont {Carrasquilla}}, \ and\ \bibinfo {author} {\bibfnamefont {Annabelle}\ \bibnamefont {Bohrdt}},\ }\bibfield  {title} {\enquote {\bibinfo {title} {Neural network approach to quasiparticle dispersions in doped antiferromagnets},}\ }\href@noop {} {\bibfield  {journal} {\bibinfo  {journal} {Communications Physics}\ }\textbf {\bibinfo {volume} {7}},\ \bibinfo {pages} {187} (\bibinfo {year} {2024})}\BibitemShut {NoStop}%
\bibitem [{\citenamefont {Sinibaldi}\ \emph {et~al.}(2024)\citenamefont {Sinibaldi}, \citenamefont {Hendry}, \citenamefont {Vicentini},\ and\ \citenamefont {Carleo}}]{sinibaldi2024timedependentneuralgalerkinmethod}%
  \BibitemOpen
  \bibfield  {author} {\bibinfo {author} {\bibfnamefont {Alessandro}\ \bibnamefont {Sinibaldi}}, \bibinfo {author} {\bibfnamefont {Douglas}\ \bibnamefont {Hendry}}, \bibinfo {author} {\bibfnamefont {Filippo}\ \bibnamefont {Vicentini}}, \ and\ \bibinfo {author} {\bibfnamefont {Giuseppe}\ \bibnamefont {Carleo}},\ }\href {https://arxiv.org/abs/2412.11778} {\enquote {\bibinfo {title} {Time-dependent neural galerkin method for quantum dynamics},}\ } (\bibinfo {year} {2024}),\ \Eprint {http://arxiv.org/abs/2412.11778} {arXiv:2412.11778 [quant-ph]} \BibitemShut {NoStop}%
\bibitem [{\citenamefont {de~Walle}\ \emph {et~al.}(2024)\citenamefont {de~Walle}, \citenamefont {Schmitt},\ and\ \citenamefont {Bohrdt}}]{vandewalle2024manybodydynamicsexplicitlytimedependent}%
  \BibitemOpen
  \bibfield  {author} {\bibinfo {author} {\bibfnamefont {Anka~Van}\ \bibnamefont {de~Walle}}, \bibinfo {author} {\bibfnamefont {Markus}\ \bibnamefont {Schmitt}}, \ and\ \bibinfo {author} {\bibfnamefont {Annabelle}\ \bibnamefont {Bohrdt}},\ }\href {https://arxiv.org/abs/2412.11830} {\enquote {\bibinfo {title} {Many-body dynamics with explicitly time-dependent neural quantum states},}\ } (\bibinfo {year} {2024}),\ \Eprint {http://arxiv.org/abs/2412.11830} {arXiv:2412.11830 [quant-ph]} \BibitemShut {NoStop}%
\bibitem [{\citenamefont {Semeghini}\ \emph {et~al.}(2021)\citenamefont {Semeghini}, \citenamefont {Levine}, \citenamefont {Keesling}, \citenamefont {Ebadi}, \citenamefont {Wang}, \citenamefont {Bluvstein}, \citenamefont {Verresen}, \citenamefont {Pichler}, \citenamefont {Kalinowski}, \citenamefont {Samajdar}, \citenamefont {Omran}, \citenamefont {Sachdev}, \citenamefont {Vishwanath}, \citenamefont {Greiner}, \citenamefont {Vuletić},\ and\ \citenamefont {Lukin}}]{RydbergSimulator2021}%
  \BibitemOpen
  \bibfield  {author} {\bibinfo {author} {\bibfnamefont {G.}~\bibnamefont {Semeghini}}, \bibinfo {author} {\bibfnamefont {H.}~\bibnamefont {Levine}}, \bibinfo {author} {\bibfnamefont {A.}~\bibnamefont {Keesling}}, \bibinfo {author} {\bibfnamefont {S.}~\bibnamefont {Ebadi}}, \bibinfo {author} {\bibfnamefont {T.~T.}\ \bibnamefont {Wang}}, \bibinfo {author} {\bibfnamefont {D.}~\bibnamefont {Bluvstein}}, \bibinfo {author} {\bibfnamefont {R.}~\bibnamefont {Verresen}}, \bibinfo {author} {\bibfnamefont {H.}~\bibnamefont {Pichler}}, \bibinfo {author} {\bibfnamefont {M.}~\bibnamefont {Kalinowski}}, \bibinfo {author} {\bibfnamefont {R.}~\bibnamefont {Samajdar}}, \bibinfo {author} {\bibfnamefont {A.}~\bibnamefont {Omran}}, \bibinfo {author} {\bibfnamefont {S.}~\bibnamefont {Sachdev}}, \bibinfo {author} {\bibfnamefont {A.}~\bibnamefont {Vishwanath}}, \bibinfo {author} {\bibfnamefont {M.}~\bibnamefont {Greiner}}, \bibinfo {author} {\bibfnamefont {V.}~\bibnamefont {Vuletić}}, \ and\ \bibinfo {author} {\bibfnamefont
  {M.~D.}\ \bibnamefont {Lukin}},\ }\bibfield  {title} {\enquote {\bibinfo {title} {Probing topological spin liquids on a programmable quantum simulator},}\ }\href {\doibase 10.1126/science.abi8794} {\bibfield  {journal} {\bibinfo  {journal} {Science}\ }\textbf {\bibinfo {volume} {374}},\ \bibinfo {pages} {1242–1247} (\bibinfo {year} {2021})}\BibitemShut {NoStop}%
\bibitem [{\citenamefont {Giudici}\ \emph {et~al.}(2022)\citenamefont {Giudici}, \citenamefont {Lukin},\ and\ \citenamefont {Pichler}}]{giudici2022dynamical}%
  \BibitemOpen
  \bibfield  {author} {\bibinfo {author} {\bibfnamefont {Giuliano}\ \bibnamefont {Giudici}}, \bibinfo {author} {\bibfnamefont {Mikhail~D}\ \bibnamefont {Lukin}}, \ and\ \bibinfo {author} {\bibfnamefont {Hannes}\ \bibnamefont {Pichler}},\ }\href@noop {} {\enquote {\bibinfo {title} {Dynamical preparation of quantum spin liquids in rydberg atom arrays},}\ } (\bibinfo {year} {2022}),\ \Eprint {http://arxiv.org/abs/2202.09372} {arXiv:2202.09372 [quant-ph]} \BibitemShut {NoStop}%
\bibitem [{\citenamefont {Carrasquilla}\ \emph {et~al.}(2019)\citenamefont {Carrasquilla}, \citenamefont {Torlai}, \citenamefont {Melko},\ and\ \citenamefont {Aolita}}]{Carrasquilla2019}%
  \BibitemOpen
  \bibfield  {author} {\bibinfo {author} {\bibfnamefont {Juan}\ \bibnamefont {Carrasquilla}}, \bibinfo {author} {\bibfnamefont {Giacomo}\ \bibnamefont {Torlai}}, \bibinfo {author} {\bibfnamefont {Roger~G.}\ \bibnamefont {Melko}}, \ and\ \bibinfo {author} {\bibfnamefont {Leandro}\ \bibnamefont {Aolita}},\ }\bibfield  {title} {\enquote {\bibinfo {title} {Reconstructing quantum states with generative models},}\ }\href {\doibase 10.1038/s42256-019-0028-1} {\bibfield  {journal} {\bibinfo  {journal} {Nature Machine Intelligence}\ }\textbf {\bibinfo {volume} {1}},\ \bibinfo {pages} {155–161} (\bibinfo {year} {2019})}\BibitemShut {NoStop}%
\bibitem [{\citenamefont {Bennewitz}\ \emph {et~al.}(2021)\citenamefont {Bennewitz}, \citenamefont {Hopfmueller}, \citenamefont {Kulchytskyy}, \citenamefont {Carrasquilla},\ and\ \citenamefont {Ronagh}}]{Bennewitz2021NeuralEM}%
  \BibitemOpen
  \bibfield  {author} {\bibinfo {author} {\bibfnamefont {Elizabeth~R.}\ \bibnamefont {Bennewitz}}, \bibinfo {author} {\bibfnamefont {Florian}\ \bibnamefont {Hopfmueller}}, \bibinfo {author} {\bibfnamefont {Bohdan}\ \bibnamefont {Kulchytskyy}}, \bibinfo {author} {\bibfnamefont {Juan~Felipe}\ \bibnamefont {Carrasquilla}}, \ and\ \bibinfo {author} {\bibfnamefont {Pooya}\ \bibnamefont {Ronagh}},\ }\href {https://www.arxiv.com/abs/2105.08086} {\enquote {\bibinfo {title} {Neural error mitigation of near-term quantum simulations},}\ } (\bibinfo {year} {2021}),\ \Eprint {http://arxiv.org/abs/2105.08086} {2105.08086} \BibitemShut {NoStop}%
\bibitem [{\citenamefont {Lange}\ \emph {et~al.}(2025)\citenamefont {Lange}, \citenamefont {Bornet}, \citenamefont {Emperauger}, \citenamefont {Chen}, \citenamefont {Lahaye}, \citenamefont {Kienle}, \citenamefont {Browaeys},\ and\ \citenamefont {Bohrdt}}]{Lange_2025}%
  \BibitemOpen
  \bibfield  {author} {\bibinfo {author} {\bibfnamefont {Hannah}\ \bibnamefont {Lange}}, \bibinfo {author} {\bibfnamefont {Guillaume}\ \bibnamefont {Bornet}}, \bibinfo {author} {\bibfnamefont {Gabriel}\ \bibnamefont {Emperauger}}, \bibinfo {author} {\bibfnamefont {Cheng}\ \bibnamefont {Chen}}, \bibinfo {author} {\bibfnamefont {Thierry}\ \bibnamefont {Lahaye}}, \bibinfo {author} {\bibfnamefont {Stefan}\ \bibnamefont {Kienle}}, \bibinfo {author} {\bibfnamefont {Antoine}\ \bibnamefont {Browaeys}}, \ and\ \bibinfo {author} {\bibfnamefont {Annabelle}\ \bibnamefont {Bohrdt}},\ }\bibfield  {title} {\enquote {\bibinfo {title} {Transformer neural networks and quantum simulators: a hybrid approach for simulating strongly correlated systems},}\ }\href {\doibase 10.22331/q-2025-03-26-1675} {\bibfield  {journal} {\bibinfo  {journal} {Quantum}\ }\textbf {\bibinfo {volume} {9}},\ \bibinfo {pages} {1675} (\bibinfo {year} {2025})}\BibitemShut {NoStop}%
\bibitem [{\citenamefont {Carleo}\ and\ \citenamefont {Troyer}(2017)}]{Carleo2017}%
  \BibitemOpen
  \bibfield  {author} {\bibinfo {author} {\bibfnamefont {Giuseppe}\ \bibnamefont {Carleo}}\ and\ \bibinfo {author} {\bibfnamefont {Matthias}\ \bibnamefont {Troyer}},\ }\bibfield  {title} {\enquote {\bibinfo {title} {Solving the quantum many-body problem with artificial neural networks},}\ }\href {\doibase 10.1126/science.aag2302} {\bibfield  {journal} {\bibinfo  {journal} {Science}\ }\textbf {\bibinfo {volume} {355}},\ \bibinfo {pages} {602–606} (\bibinfo {year} {2017})}\BibitemShut {NoStop}%
\bibitem [{\citenamefont {Abadi}\ \emph {et~al.}(2015)\citenamefont {Abadi}, \citenamefont {Agarwal}, \citenamefont {Barham}, \citenamefont {Brevdo}, \citenamefont {Chen}, \citenamefont {Citro}, \citenamefont {Corrado}, \citenamefont {Davis}, \citenamefont {Dean}, \citenamefont {Devin}, \citenamefont {Ghemawat}, \citenamefont {Goodfellow}, \citenamefont {Harp}, \citenamefont {Irving}, \citenamefont {Isard}, \citenamefont {Jia}, \citenamefont {Jozefowicz}, \citenamefont {Kaiser}, \citenamefont {Kudlur}, \citenamefont {Levenberg}, \citenamefont {Man\'{e}}, \citenamefont {Monga}, \citenamefont {Moore}, \citenamefont {Murray}, \citenamefont {Olah}, \citenamefont {Schuster}, \citenamefont {Shlens}, \citenamefont {Steiner}, \citenamefont {Sutskever}, \citenamefont {Talwar}, \citenamefont {Tucker}, \citenamefont {Vanhoucke}, \citenamefont {Vasudevan}, \citenamefont {Vi\'{e}gas}, \citenamefont {Vinyals}, \citenamefont {Warden}, \citenamefont {Wattenberg}, \citenamefont {Wicke}, \citenamefont {Yu},\ and\ \citenamefont
  {Zheng}}]{tensorflow2015-whitepaper}%
  \BibitemOpen
  \bibfield  {author} {\bibinfo {author} {\bibfnamefont {Mart\'{\i}n}\ \bibnamefont {Abadi}}, \bibinfo {author} {\bibfnamefont {Ashish}\ \bibnamefont {Agarwal}}, \bibinfo {author} {\bibfnamefont {Paul}\ \bibnamefont {Barham}}, \bibinfo {author} {\bibfnamefont {Eugene}\ \bibnamefont {Brevdo}}, \bibinfo {author} {\bibfnamefont {Zhifeng}\ \bibnamefont {Chen}}, \bibinfo {author} {\bibfnamefont {Craig}\ \bibnamefont {Citro}}, \bibinfo {author} {\bibfnamefont {Greg~S.}\ \bibnamefont {Corrado}}, \bibinfo {author} {\bibfnamefont {Andy}\ \bibnamefont {Davis}}, \bibinfo {author} {\bibfnamefont {Jeffrey}\ \bibnamefont {Dean}}, \bibinfo {author} {\bibfnamefont {Matthieu}\ \bibnamefont {Devin}}, \bibinfo {author} {\bibfnamefont {Sanjay}\ \bibnamefont {Ghemawat}}, \bibinfo {author} {\bibfnamefont {Ian}\ \bibnamefont {Goodfellow}}, \bibinfo {author} {\bibfnamefont {Andrew}\ \bibnamefont {Harp}}, \bibinfo {author} {\bibfnamefont {Geoffrey}\ \bibnamefont {Irving}}, \bibinfo {author} {\bibfnamefont {Michael}\ \bibnamefont
  {Isard}}, \bibinfo {author} {\bibfnamefont {Yangqing}\ \bibnamefont {Jia}}, \bibinfo {author} {\bibfnamefont {Rafal}\ \bibnamefont {Jozefowicz}}, \bibinfo {author} {\bibfnamefont {Lukasz}\ \bibnamefont {Kaiser}}, \bibinfo {author} {\bibfnamefont {Manjunath}\ \bibnamefont {Kudlur}}, \bibinfo {author} {\bibfnamefont {Josh}\ \bibnamefont {Levenberg}}, \bibinfo {author} {\bibfnamefont {Dandelion}\ \bibnamefont {Man\'{e}}}, \bibinfo {author} {\bibfnamefont {Rajat}\ \bibnamefont {Monga}}, \bibinfo {author} {\bibfnamefont {Sherry}\ \bibnamefont {Moore}}, \bibinfo {author} {\bibfnamefont {Derek}\ \bibnamefont {Murray}}, \bibinfo {author} {\bibfnamefont {Chris}\ \bibnamefont {Olah}}, \bibinfo {author} {\bibfnamefont {Mike}\ \bibnamefont {Schuster}}, \bibinfo {author} {\bibfnamefont {Jonathon}\ \bibnamefont {Shlens}}, \bibinfo {author} {\bibfnamefont {Benoit}\ \bibnamefont {Steiner}}, \bibinfo {author} {\bibfnamefont {Ilya}\ \bibnamefont {Sutskever}}, \bibinfo {author} {\bibfnamefont {Kunal}\ \bibnamefont {Talwar}},
  \bibinfo {author} {\bibfnamefont {Paul}\ \bibnamefont {Tucker}}, \bibinfo {author} {\bibfnamefont {Vincent}\ \bibnamefont {Vanhoucke}}, \bibinfo {author} {\bibfnamefont {Vijay}\ \bibnamefont {Vasudevan}}, \bibinfo {author} {\bibfnamefont {Fernanda}\ \bibnamefont {Vi\'{e}gas}}, \bibinfo {author} {\bibfnamefont {Oriol}\ \bibnamefont {Vinyals}}, \bibinfo {author} {\bibfnamefont {Pete}\ \bibnamefont {Warden}}, \bibinfo {author} {\bibfnamefont {Martin}\ \bibnamefont {Wattenberg}}, \bibinfo {author} {\bibfnamefont {Martin}\ \bibnamefont {Wicke}}, \bibinfo {author} {\bibfnamefont {Yuan}\ \bibnamefont {Yu}}, \ and\ \bibinfo {author} {\bibfnamefont {Xiaoqiang}\ \bibnamefont {Zheng}},\ }\href {https://www.tensorflow.org/} {\enquote {\bibinfo {title} {{TensorFlow}: Large-scale machine learning on heterogeneous systems},}\ } (\bibinfo {year} {2015}),\ \bibinfo {note} {software available from tensorflow.org}\BibitemShut {NoStop}%
\bibitem [{\citenamefont {Harris}\ \emph {et~al.}(2020)\citenamefont {Harris}, \citenamefont {Millman}, \citenamefont {van~der Walt}, \citenamefont {Gommers}, \citenamefont {Virtanen}, \citenamefont {Cournapeau}, \citenamefont {Wieser}, \citenamefont {Taylor}, \citenamefont {Berg}, \citenamefont {Smith}, \citenamefont {Kern}, \citenamefont {Picus}, \citenamefont {Hoyer}, \citenamefont {van Kerkwijk}, \citenamefont {Brett}, \citenamefont {Haldane}, \citenamefont {del R{\'i}o}, \citenamefont {Wiebe}, \citenamefont {Peterson}, \citenamefont {G{\'e}rard-Marchant}, \citenamefont {Sheppard}, \citenamefont {Reddy}, \citenamefont {Weckesser}, \citenamefont {Abbasi}, \citenamefont {Gohlke},\ and\ \citenamefont {Oliphant}}]{Harris2020}%
  \BibitemOpen
  \bibfield  {author} {\bibinfo {author} {\bibfnamefont {Charles~R.}\ \bibnamefont {Harris}}, \bibinfo {author} {\bibfnamefont {K.~Jarrod}\ \bibnamefont {Millman}}, \bibinfo {author} {\bibfnamefont {St{\'e}fan~J.}\ \bibnamefont {van~der Walt}}, \bibinfo {author} {\bibfnamefont {Ralf}\ \bibnamefont {Gommers}}, \bibinfo {author} {\bibfnamefont {Pauli}\ \bibnamefont {Virtanen}}, \bibinfo {author} {\bibfnamefont {David}\ \bibnamefont {Cournapeau}}, \bibinfo {author} {\bibfnamefont {Eric}\ \bibnamefont {Wieser}}, \bibinfo {author} {\bibfnamefont {Julian}\ \bibnamefont {Taylor}}, \bibinfo {author} {\bibfnamefont {Sebastian}\ \bibnamefont {Berg}}, \bibinfo {author} {\bibfnamefont {Nathaniel~J.}\ \bibnamefont {Smith}}, \bibinfo {author} {\bibfnamefont {Robert}\ \bibnamefont {Kern}}, \bibinfo {author} {\bibfnamefont {Matti}\ \bibnamefont {Picus}}, \bibinfo {author} {\bibfnamefont {Stephan}\ \bibnamefont {Hoyer}}, \bibinfo {author} {\bibfnamefont {Marten~H.}\ \bibnamefont {van Kerkwijk}}, \bibinfo {author}
  {\bibfnamefont {Matthew}\ \bibnamefont {Brett}}, \bibinfo {author} {\bibfnamefont {Allan}\ \bibnamefont {Haldane}}, \bibinfo {author} {\bibfnamefont {Jaime~Fern{\'a}ndez}\ \bibnamefont {del R{\'i}o}}, \bibinfo {author} {\bibfnamefont {Mark}\ \bibnamefont {Wiebe}}, \bibinfo {author} {\bibfnamefont {Pearu}\ \bibnamefont {Peterson}}, \bibinfo {author} {\bibfnamefont {Pierre}\ \bibnamefont {G{\'e}rard-Marchant}}, \bibinfo {author} {\bibfnamefont {Kevin}\ \bibnamefont {Sheppard}}, \bibinfo {author} {\bibfnamefont {Tyler}\ \bibnamefont {Reddy}}, \bibinfo {author} {\bibfnamefont {Warren}\ \bibnamefont {Weckesser}}, \bibinfo {author} {\bibfnamefont {Hameer}\ \bibnamefont {Abbasi}}, \bibinfo {author} {\bibfnamefont {Christoph}\ \bibnamefont {Gohlke}}, \ and\ \bibinfo {author} {\bibfnamefont {Travis~E.}\ \bibnamefont {Oliphant}},\ }\bibfield  {title} {\enquote {\bibinfo {title} {Array programming with numpy},}\ }\href {\doibase 10.1038/s41586-020-2649-2} {\bibfield  {journal} {\bibinfo  {journal} {Nature}\ }\textbf
  {\bibinfo {volume} {585}},\ \bibinfo {pages} {357--362} (\bibinfo {year} {2020})}\BibitemShut {NoStop}%
\bibitem [{\citenamefont {Zhou}\ \emph {et~al.}(2016)\citenamefont {Zhou}, \citenamefont {Wu}, \citenamefont {Zhang},\ and\ \citenamefont {Zhou}}]{zhou2016minimal}%
  \BibitemOpen
  \bibfield  {author} {\bibinfo {author} {\bibfnamefont {Guo-Bing}\ \bibnamefont {Zhou}}, \bibinfo {author} {\bibfnamefont {Jianxin}\ \bibnamefont {Wu}}, \bibinfo {author} {\bibfnamefont {Chen-Lin}\ \bibnamefont {Zhang}}, \ and\ \bibinfo {author} {\bibfnamefont {Zhi-Hua}\ \bibnamefont {Zhou}},\ }\href@noop {} {\enquote {\bibinfo {title} {Minimal gated unit for recurrent neural networks},}\ } (\bibinfo {year} {2016}),\ \Eprint {http://arxiv.org/abs/1603.09420} {arXiv:1603.09420 [cs.NE]} \BibitemShut {NoStop}%
\bibitem [{\citenamefont {Shen}(2019)}]{shen2019mutual}%
  \BibitemOpen
  \bibfield  {author} {\bibinfo {author} {\bibfnamefont {Huitao}\ \bibnamefont {Shen}},\ }\href@noop {} {\enquote {\bibinfo {title} {Mutual information scaling and expressive power of sequence models},}\ } (\bibinfo {year} {2019}),\ \Eprint {http://arxiv.org/abs/1905.04271} {arXiv:1905.04271 [cs.LG]} \BibitemShut {NoStop}%
\bibitem [{\citenamefont {Kingma}\ and\ \citenamefont {Ba}(2015)}]{AdamPaper}%
  \BibitemOpen
  \bibfield  {author} {\bibinfo {author} {\bibfnamefont {Diederik~P.}\ \bibnamefont {Kingma}}\ and\ \bibinfo {author} {\bibfnamefont {Jimmy}\ \bibnamefont {Ba}},\ }\bibfield  {title} {\enquote {\bibinfo {title} {Adam: {A} method for stochastic optimization},}\ }in\ \href {http://arxiv.org/abs/1412.6980} {\emph {\bibinfo {booktitle} {3rd International Conference on Learning Representations, {ICLR} 2015, San Diego, CA, USA, May 7-9, 2015, Conference Track Proceedings}}},\ \bibinfo {editor} {edited by\ \bibinfo {editor} {\bibfnamefont {Yoshua}\ \bibnamefont {Bengio}}\ and\ \bibinfo {editor} {\bibfnamefont {Yann}\ \bibnamefont {LeCun}}}\ (\bibinfo {year} {2015})\BibitemShut {NoStop}%
\bibitem [{\citenamefont {Torlai}\ and\ \citenamefont {Fishman}(2020)}]{pastaq}%
  \BibitemOpen
  \bibfield  {author} {\bibinfo {author} {\bibfnamefont {Giacomo}\ \bibnamefont {Torlai}}\ and\ \bibinfo {author} {\bibfnamefont {Matthew}\ \bibnamefont {Fishman}},\ }\href {https://github.com/GTorlai/PastaQ.jl/} {\enquote {\bibinfo {title} {\mbox{PastaQ}: A package for simulation, tomography and analysis of quantum computers},}\ } (\bibinfo {year} {2020})\BibitemShut {NoStop}%
\bibitem [{\citenamefont {Fishman}\ \emph {et~al.}(2022)\citenamefont {Fishman}, \citenamefont {White},\ and\ \citenamefont {Stoudenmire}}]{Fishman_2022}%
  \BibitemOpen
  \bibfield  {author} {\bibinfo {author} {\bibfnamefont {Matthew}\ \bibnamefont {Fishman}}, \bibinfo {author} {\bibfnamefont {Steven}\ \bibnamefont {White}}, \ and\ \bibinfo {author} {\bibfnamefont {Edwin}\ \bibnamefont {Stoudenmire}},\ }\bibfield  {title} {\enquote {\bibinfo {title} {The itensor software library for tensor network calculations},}\ }\href {\doibase 10.21468/scipostphyscodeb.4} {\bibfield  {journal} {\bibinfo  {journal} {SciPost Physics Codebases}\ } (\bibinfo {year} {2022}),\ 10.21468/scipostphyscodeb.4}\BibitemShut {NoStop}%
\end{thebibliography}%

\end{document}